\documentclass[lettersize,journal]{IEEEtran}
\usepackage{amsmath,amsfonts}
\usepackage{algorithmic}
\usepackage{algorithm}
\usepackage{array}
\usepackage[caption=false,font=normalsize,labelfont=sf,textfont=sf]{subfig}
\usepackage{textcomp}
\usepackage{stfloats}
\usepackage{url}
\usepackage{verbatim}
\usepackage{graphicx}
\usepackage{cite}
\usepackage{lscape}
\usepackage{makecell}
\usepackage{wasysym}
\usepackage{pifont}
\usepackage{fontawesome}
\usepackage{utfsym}
\usepackage{multirow}
\usepackage{multicol}
\usepackage{tikz}
\usepackage{longtable}
\usepackage{booktabs}
\usetikzlibrary{patterns}
\usetikzlibrary{patterns.meta}
\usepackage{scalerel}
\usepackage{tikz}
\usetikzlibrary{svg.path}
\usepackage{hyperref}

\definecolor{orcidlogocol}{HTML}{A6CE39}
\tikzset{
  orcidlogo/.pic={
    \fill[orcidlogocol] svg{M256,128c0,70.7-57.3,128-128,128C57.3,256,0,198.7,0,128C0,57.3,57.3,0,128,0C198.7,0,256,57.3,256,128z};
    \fill[white] svg{M86.3,186.2H70.9V79.1h15.4v48.4V186.2z}
                 svg{M108.9,79.1h41.6c39.6,0,57,28.3,57,53.6c0,27.5-21.5,53.6-56.8,53.6h-41.8V79.1z M124.3,172.4h24.5c34.9,0,42.9-26.5,42.9-39.7c0-21.5-13.7-39.7-43.7-39.7h-23.7V172.4z}
                 svg{M88.7,56.8c0,5.5-4.5,10.1-10.1,10.1c-5.6,0-10.1-4.6-10.1-10.1c0-5.6,4.5-10.1,10.1-10.1C84.2,46.7,88.7,51.3,88.7,56.8z};
  }
}

\newcommand\orcidicon[1]{\href{https://orcid.org/#1}{\mbox{\scalerel*{
\begin{tikzpicture}[yscale=-1,transform shape]
\pic{orcidlogo};
\end{tikzpicture}
}{|}}}}

\newcommand{\lefthalfcircle}{%
\begin{tikzpicture}
    \draw (0,0) circle (0.08cm);
    \path[pattern={Lines[angle=45,distance=1pt]}, pattern color=black!90]
    (0,0) -- (0, 0.08) arc(90:270:0.085) -- cycle;
\end{tikzpicture}}
\newcommand{\righthalfcircle}{%
\begin{tikzpicture}
    \draw (0,0) circle (0.08cm);
    \path[pattern={Lines[angle=45,distance=1pt]}, pattern color=black!90]
    (0,0) circle (0.08cm);
\end{tikzpicture}}
\begin{document}

\title{Maritime Cybersecurity: A Comprehensive Review}
\author{Meixuan Li \orcidicon{0009-0004-8515-2659}\, Jianying Zhou \orcidicon{0000-0003-0594-0432}\, Sudipta Chattopadhyay \orcidicon{0000-0002-4843-5391}\, Mark Goh \orcidicon{0000-0002-8282-6628}
\thanks{Manuscript received 01 November 2024; revised TBD; accepted TBD. Date of publication TBD; date of current version 01 November 2024}}

\markboth{IEEE TRANSACTIONS ON INTELLIGENT TRANSPORTATION SYSTEMS}
{Shell \MakeLowercase{\textit{et al.}}: A Sample Article Using IEEEtran.cls for IEEE Journals}


\maketitle

\begin{abstract}
The maritime industry stands at a critical juncture, where the imperative for technological advancement intersects with the pressing need for robust cybersecurity measures. Maritime cybersecurity refers to the protection of computer systems and digital assests within the maritime industry, as well as the broader network of interconnected components that make up the maritime ecosystem. In this survey, we aim to identify the significant domains of maritime cybersecurity and measure their effectiveness. We have provided an in-depth analysis of threats in key maritime systems, including AIS, GNSS, ECDIS, VDR, RADAR, VSAT, and GMDSS, while exploring real-world cyber incidents that have impacted the sector. A multi-dimensional taxonomy of maritime cyber attacks is presented, offering insights into threat actors, motivations, and impacts. We have also evaluated various security solutions, from integrated solutions to component specific solutions. Finally, we have shared open challenges and future solutions. By addressing all these critical issues with key interconnected aspects, this review aims to foster a more resilient maritime ecosystem.
\end{abstract}


\section{Introduction}
\begin{figure}[h]
  \centering
  
  \includegraphics[width=8.5cm]{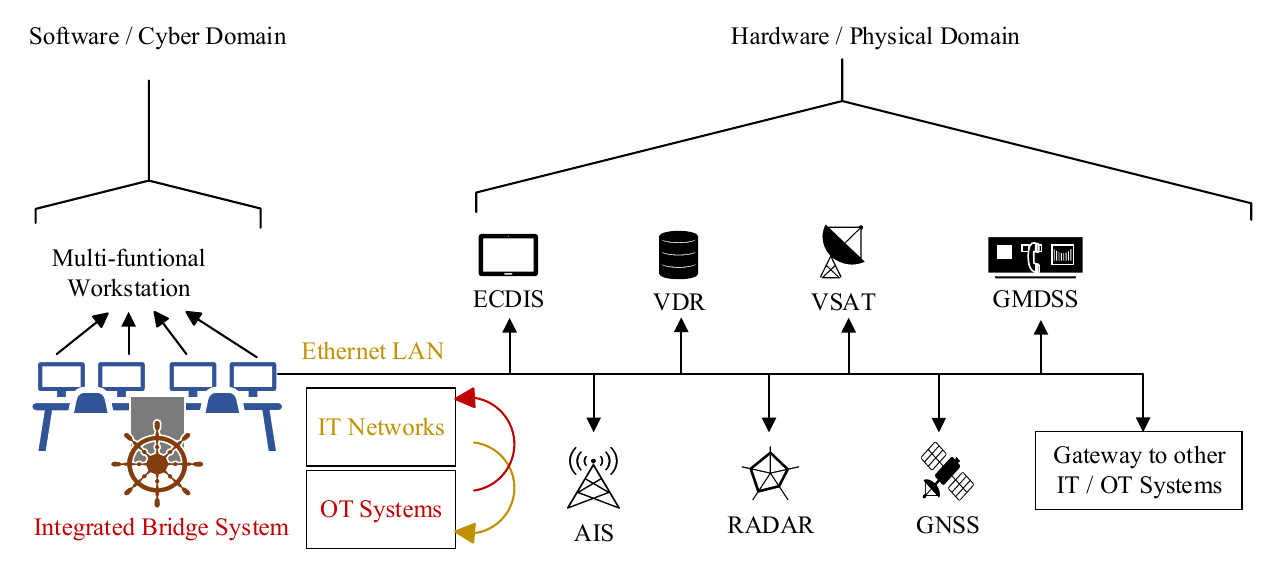}
  \caption{Illustration of a typical maritime shipboard IT/OT system that combines hardware and software to monitor and control physical devices in real time. The increasing interconnectivity of IT and OT systems have expanded the potential attack surface.}
  \label{fig:OTsystem}
\end{figure}

\IEEEPARstart{T}{he} global maritime digital technology industry is estimated to be worth \$345 billion by 2030, up from a previous forecast of \$279 billion \cite{digital}. Over the past few years, the maritime industry has progressed rapidly and its evolution has extended into several domains. From increased use of digital systems \cite{gavalas2022digital}, enhanced satellite communication \cite{alqurashi2022maritime}, and Internet-of-Things (IoT)-enabled port infrastructure \cite{cil2022internet} to development tendencies of autonomous shipping \cite{karetnikov2019technology}, a series of studies have been conducted with a focus on the advancements of maritime technology. Despite this progress, maritime cybersecurity has been inadequately addressed. Specifically, the maritime industry is at risk of becoming a target for a series of cyber threats and attack vectors that arise from the interconnection of diverse technical tools. This includes, among others, IoT networks and telecommunications. The maritime cyber environment, as described by Sotiria \cite{lagouvardou2018maritime}, encompasses the networked systems of both Information Technology (IT) and cyber-physical systems (i.e.,  Operational Technology (OT)). It includes communication networks that allow data to flow from IT systems of a ship to the OT systems. Such is accomplished via programmable logic controllers (PLCs) and various sensors and ultimately controls advanced navigation tools like Global Positioning Systems (GPS). The integration of IT and OT in the maritime sector introduces heightened cybersecurity risks, as traditionally isolated OT systems become more susceptible to cyber attacks through their connections to IT networks (See Figure~\ref{fig:OTsystem}). As the range of potential threats broaden, more vulnerabilities and entry points have been discovered in maritime systems.

Vessels have become vulnerable to cyberattacks originating from both internally and externally. With the rising levels of connectivity, the scope of maritime cybersecurity extends beyond the vessels and is characterized by complex interactions among stakeholders and components, such as shippers and ports. Unlike the automotive industry, where the impact of cyberattacks tend to be more localized, affecting single vehicles or restricted regions, the potential impact of a cyberattack on a large vessel could destabilize global supply chains and inflict billions in economic losses \cite{roberts2019combined}. Moreover, the risk extends beyond individual ships - a cyberattack targeting shared infrastructure like a VSAT service supplier could trigger a fleetwide or cross-fleet event \cite{flint}, magnifying the damage across multiple vessels and shipping companies simultaneously.
\begin{table*}[h]\scriptsize
  \caption{Positioning of our study with respect to prior surveys on maritime cybersecurity}
  \label{tab:new compare}
  \centering
  \begin{tabular}{p{1.8cm} p{3cm}<{\centering} p{2cm}<{\centering}p{2cm}<{\centering}p{1.8cm}<{\centering}
 p{2cm}<{\centering} p{2cm}<{\centering}  }
    \toprule
    Survey 
    &Component-Based Approach to Maritime Threats Overview
    &Discussing Broader Maritime Security Issues
    &Analyzing Maritime Attack Incidents
    &Holistic Mitigation Measures
    &Component-specific Mitigation Measures
    &Including Directions \& Challenges  \\
    \midrule
    Park \textit{et al.} \cite{park2019cybersecurity} & \usym{2613} & \Circle &\LEFTcircle& \LEFTcircle &\Circle&\usym{2613} \\
    Ismail \textit{et al.} \cite{ismail2021survey}& \usym{2613} & \Circle &\Circle& \LEFTcircle&\Circle&\usym{2613}\\
    Farah \textit{et al.} \cite{ben2022cyber}& \checkmark &\Circle &\LEFTcircle& \Circle&\Circle&\usym{2613}\\
    Ashraf \textit{et al.} \cite{ashraf2022survey}& \checkmark&\Circle & \Circle &\LEFTcircle&\Circle&\usym{2613}\\
    Yu \textit{et al.} \cite{yu2023literature}&\usym{2613}&\Circle & \LEFTcircle&\LEFTcircle&\Circle&\checkmark\\
    Our survey& \checkmark&\CIRCLE& \CIRCLE&\CIRCLE&\CIRCLE&\checkmark\\
    \midrule
    \multicolumn{7}{c}{\checkmark =Yes, \usym{2613} = No, \Circle = Not Mentioned, \LEFTcircle = Room for Improvement, \CIRCLE = Thoroughly Addressed} \\
    \bottomrule
  \end{tabular}
\end{table*}
\subsection{Existing Work Comparisons}
In the last few years, several surveys on maritime cybersecurity have been 
published~\cite{ashraf2022survey,ben2022cyber,ben2022cyber,park2019cybersecurity,yu2023literature}. 
Park \textit{et al.}~\cite{park2019cybersecurity} 
adopted a high-level perspective of cybersecurity threats and risk control alternatives in the marine industry as a whole. Ismail \textit{et al.}~\cite{ismail2021survey} focused specifically on maritime cybersecurity in the Indian Ocean region with limited depth on maritime component-specific topics. Farah \textit{et al.}~\cite{ben2022cyber} provided a mapping and classification of on-vessel core equipment and in-port infrastructure and included a detailed breakdown of electro-mechanical systems, electronic systems and communication systems. 
Nevertheless, the authors \cite{ben2022cyber} did not offer comprehensive technical information on particular cybersecurity measures. In contrast to more general maritime cybersecurity, Ashraf \textit{et al.}~\cite{ashraf2022survey} offered a comprehensive classification of cyberattacks in the marine domain, with a primary focus on risks related to the Internet of Things. However, this work also did not elaborate on methods of mitigation.  Yu \textit{et al.}~\cite{yu2023literature} used bibliometric methods to provide a review of academic literature on maritime cybersecurity with limited discussion of technical details in cybersecurity solutions. Hence, there is still room to classify novel threats and countermeasures in literature that are more sophisticated and highly targeted, and our survey aims to fill this gap. 

Although security attacks against maritime IoT devices have received considerable attention during the last decade, the significance of IoT enabled attacks is not always fully assessed. We focus on \textit{verified} attacks, i.e, either real-world incidents or attacks that have been implemented and recently published by researchers including the attacks that are theoretically possible. Based on the analysis of these attacks, we summarize security solutions that can effectively mitigate such threats. Unlike  existing works that involve limited discussions on countermeasures, our study will provide a thorough review of the latest defensive strategies and technologies. Concretely, we cover not only component-specific countermeasures but also integrated solutions that can enhance the overall cybersecurity posture of maritime organizations. Table~\ref{tab:new compare} provides a comparison between our work and prior works targeting maritime cybersecurity. 

\subsection{Our Contributions}
\begin{figure}[h]
  \centering
  \includegraphics[width=9cm]{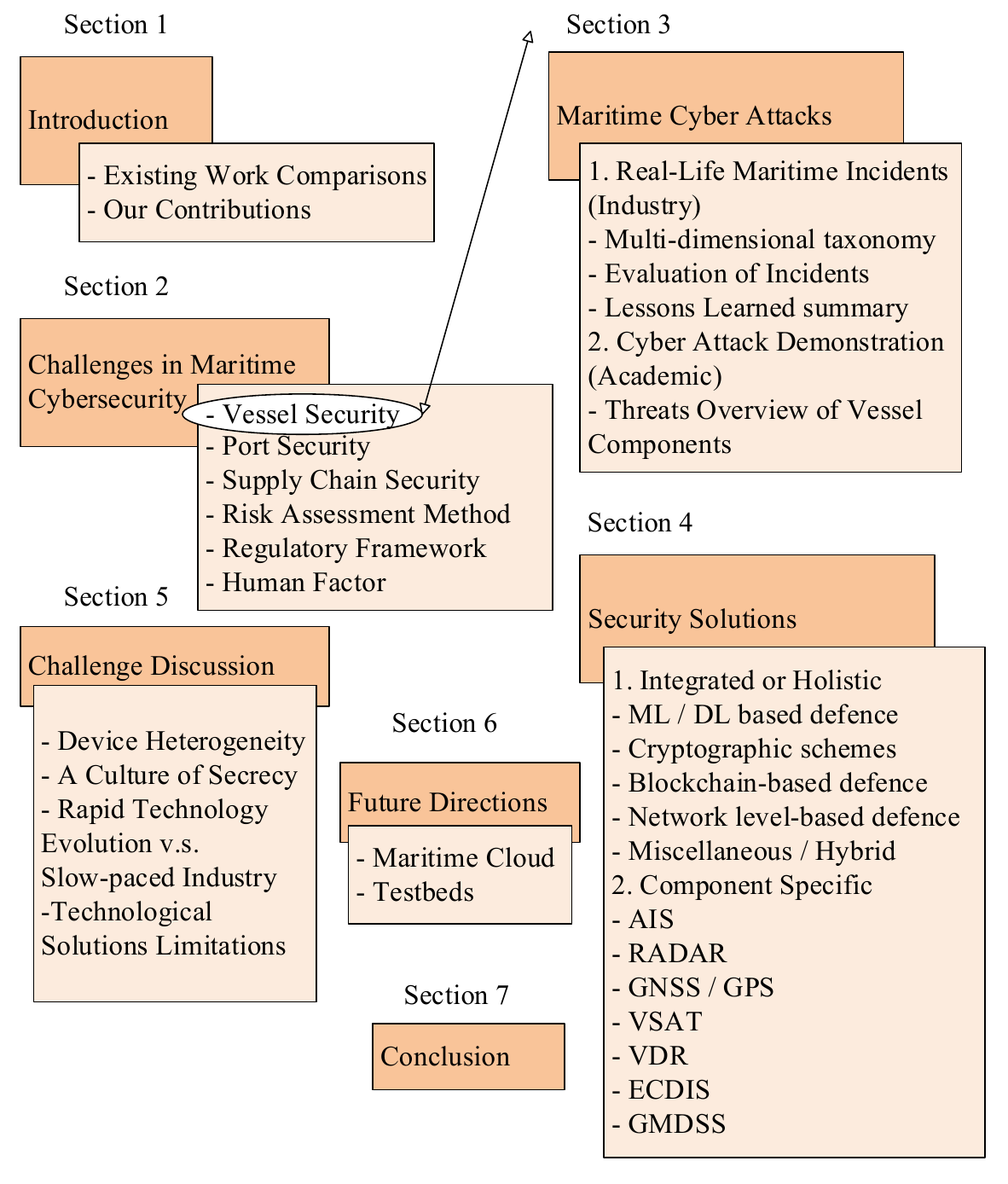}
  \caption{Structure of this study}
  \label{fig:outline}
\end{figure}

The main contributions and organizations of our study (see Figure~\ref{fig:outline}) 
are as follows: 
\begin{enumerate}
    \item We provide a broader overview of the industry's cybersecurity landscape by examining six key maritime interconnected aspects in Section~\ref{sec:components}. 

    \item Section~\ref{sec:threats a} reviews the threats of vessel components, which include RAdio Detection And Ranging (RADAR), Voyage Data Recorders (VDR), Global Navigation Satellite Systems (GNSS) / GPS, Electronic Chart Display and Information Systems (ECDIS), Automatic Identification Systems (AIS), Global Maritime Distress and Safety System (GMDSS) and Very Small Aperture Terminal (VSAT). 

    \item We illustrate and discuss a taxonomy that categorizes maritime incidents over a 10-year period, from 2014 to 2023, targeting various attack surfaces and systems of vessel in Section~\ref{sec:threats b}. 

     \item We provide a review of diverse mitigation techniques aimed at addressing security concerns in Section~\ref{sec:mitigation a} and Section~\ref{sec:mitigation b}. 

     \item After we discuss challenges and potential future directions for maritime cybersecurity in Sections~\ref{sec:challenges}-\ref{sec:direction}, we conclude 
     our article in Section~\ref{sec:conclusion}. 
\end{enumerate}

\section{Challenges in Maritime Cybersecurity}
\label{sec:components}
\begin{figure}[h]
  \centering
  \includegraphics[width=9cm]{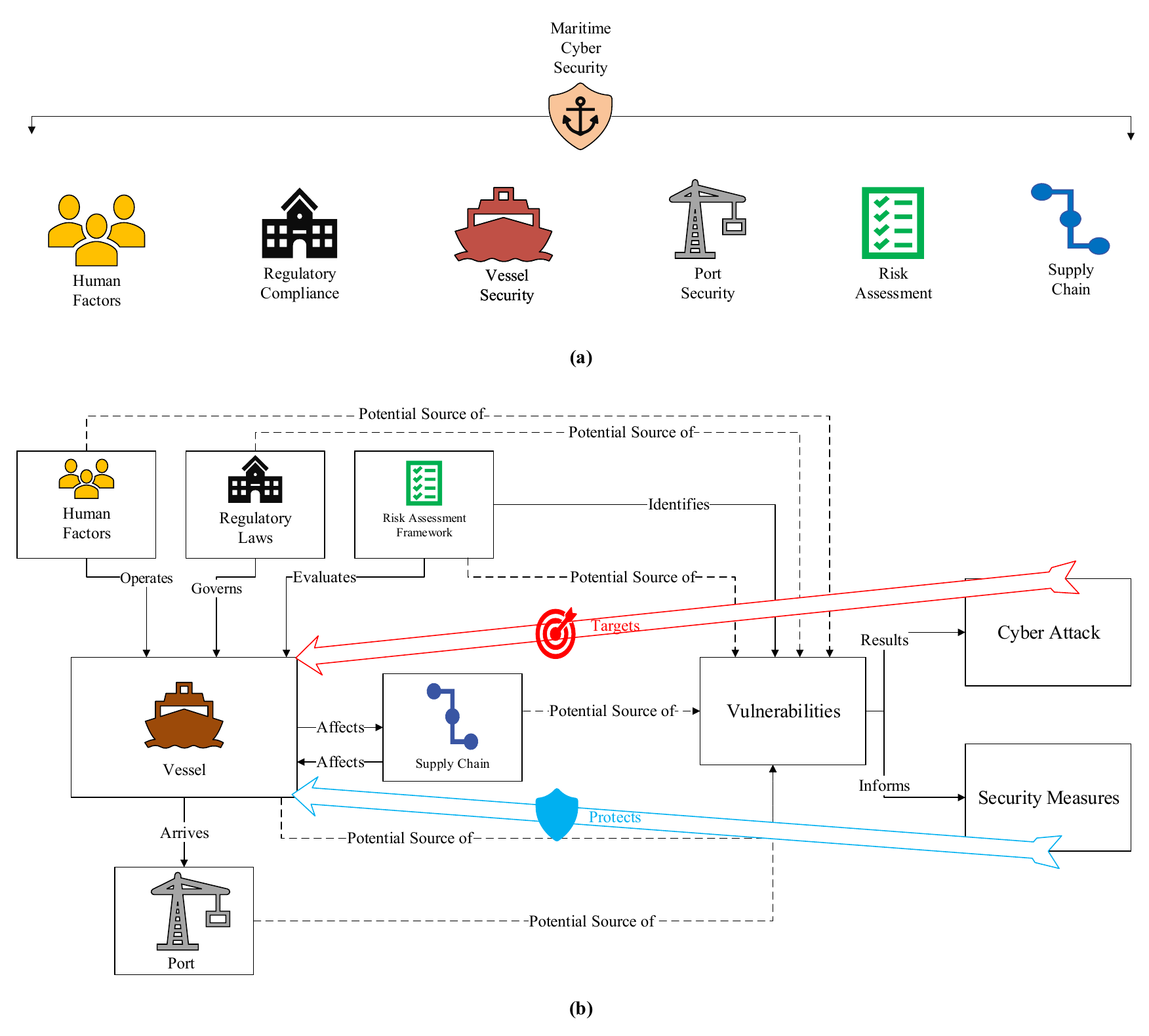}
  \caption{Six challenges in maritime cybersecurity (a) and the connection between them (b)}
  \label{fig:overview}
\end{figure}

As technology continues to advance and integrate into every aspect of maritime operations, the sector must confront vulnerabilities across multiple fronts, as Figure ~\ref{fig:overview} (a) illustrates. According to Figure~\ref{fig:overview} (b), at the center is the vessel itself, surrounded by interconnected elements that influence its security posture. Human factors, including crew actions and decisions, directly impact the vessel's operation and can inadvertently introduce risks. Regulatory laws govern vessel operations, setting standards that can either mitigate or, if outdated, potentially exacerbate cybersecurity threats. Risk assessment frameworks serve to evaluate vessel systematically. If these frameworks are not effective, they may fail to identify critical risks, leaving them hidden and unaddressed. Ports, as critical interfaces between sea and land, present cyber risks suring vessel calls. The broader supply chain both affects and is affected by vessel security, creating a reciprocal relationship of potential vulnerabilities. More details are to be introduced at the subsequent respective sections. 

\subsection{Vessel Security}
Vessel security forms the cornerstone of maritime cybersecurity, interacting with and influencing other critical aspects of the maritime network. Crew members are on the front lines of implementing security measures and are often the first to detect and respond to security threats. Well-trained crew members who are aware of security protocols and potential threats can significantly enhance a vessel's security posture \cite{berg2013impact}. However, human factors can introduce vulnerabilities (e.g. neglect security procedures). Vessel security is also governed by a complex web of international laws, conventions, and industry standards which set out security-related requirements for ships, ports, and government agencies. Although most marine incidents resulted from failing to follow international regulations related to maritime safety \cite{abuelenin2017impact}, compliance with these regulatory frameworks should be viewed as a minimum standard rather than the ultimate goal of security efforts. When a vessel enters a port, it becomes subject to the port's security procedures and infrastructure. The port security management measures can refer to the subsequent port security section. The increasing digitization of port operations, including the use of automated systems for cargo handling and vessel traffic management, creates new cybersecurity considerations for vessels \cite{inkinen2019port}. Risk assessment methodology aims help systematically identify potential security threats to a vessel and includes the protection of human life and health \cite{papageorgiou2024using}. Based on the identified risks, these methods guide the development of appropriate security measures and contingency plans. In global supply chains, transit times are heavily influenced by vessel speed and routes. Vessel operations are generally under significant time pressure, aligning with the "just-in-time" philosophy underlying most modern supply chains \cite{lind2021improving}. This pressure can cause hasty security decisions, which could jeopardize vessel safety by emphasizing temporary solutions over long-term security solutions. Since many vessel components are integrated systems made by multiple companies that operate in different industries, producers must have a deep awareness of the hazards associated with the supply chain \cite{cho2022cybersecurity}. 

This study examines the security of a particular part of the Integrated Bridge System (IBS), which could be an entry point of attack for ships. The security of the IBS is crucial as it represents a central point of control for the vessel. A vast network of systems and components is used by ships to function well in the challenging marine environment. These include navigation systems, communication equipment, and various sensors and actuators distributed throughout the ship. Better efficiency and operational capabilities are made possible by this integration, but there is a risk as well: if one component is hacked, it may give the attacker access to all other systems \cite{goodrum2018understanding}. Notably, Section 4 of this document delves into specific attacks targeting these systems and components with the potential consequences of such attacks.
\subsection{Port Security}
With forecasts showing a strong compound annual growth rate of 7.3\% from 2023 to 2031, the worldwide port security market is expected to increase significantly and could reach an estimated value of \$173.59 billion by 2031 \cite{linkedin}. This trajectory reflects the continued development of security solutions to counter new threats in marine commerce, as well as the growing acknowledgment of port security as an essential part of the infrastructure of global trade. When port security is applied comprehensively, it includes a wide range of policies, practices, and systems that are put in place to protect port operations, infrastructure, and surrounding areas against a variety of threats, such as theft, smuggling, terrorism, and other criminal activity \cite{sergi2021ports}. In order to combat the growing threat of combined cyber-physical attacks for ports, the future of port security is probably going to center on merging cyber and physical security measures \cite{adams2020port}. Its multifaceted nature highlights its complexity, necessitating the combination of technological, procedural, and physical protections to build a strong defense against both traditional and emerging threats. 

Numerous studies have addressed port infrastructure risk in detail \cite{kapalidis2022vulnerability}\cite{progoulakis2023digitalization}\cite{progoulakis2021cyber}. Port security management plays a vital role in reducing a wide range of risks and threats that are present in maritime environments. This strategy includes a number of essential elements to protect maritime operations and infrastructure, including emergency response plans, surveillance and monitoring, perimeter security, access control systems, and cargo screening \cite{christopher2009port}. While access control systems use sophisticated identification credentials, biometric technologies (such as fingerprint scanning and facial recognition), and screening procedures to regulate entry to sensitive areas, perimeter security uses cutting-edge fencing systems, barriers, and intrusion detection technologies to create a secure boundary around port facilities \cite{kusi2015port}. Furthermore, state-of-the-art monitoring systems for ship movements in port waters, along with non-intrusive cargo inspection methods like X-ray scanners and RFID tracking for cargo integrity, support prompt response times to security incidents \cite{christopher2009port}. The integrity, safety, and operational continuity of port infrastructures—which are crucial hubs in both national security frameworks and international trade networks—must be preserved by the integration of these security measures.

\subsection{Supply Chain Security}

Supply chain security refers to the laws, practices, and technological advancements created to protect physical items, information flows, and supply chain assets against threats such as theft, terrorism, and natural catastrophes\cite{yang2011risk}. The process of moving commodities by sea, involving shipping, port operations, and frequently interior transportation, is referred to as a maritime supply chain. This intricate system connects numerous ports spread over several nations and continents and encompasses a wide spectrum of stakeholders, including freight forwarders, shippers, carriers, and customs officials. The maritime supply chain is faced with a number of difficulties that could seriously jeopardize operational effectiveness and security. Among the many potential reasons of port interruptions, there include worker strikes, natural catastrophes, equipment malfunctions, and cyberattacks \cite{wendler2020port}. Port disruptions are considered a critical vulnerability.  
Information security threats, such as cyberattacks on logistics management systems and data breaches revealing private shipping information, pose a significant concern as well \cite{liu2018analysis}. Physical security threats to marine supply chains include piracy (in high-risk maritime zones), cargo theft, and the smuggling of illegal commodities \cite{yang2011risk}. These are problems that have not yet been properly resolved. As a result of continued climate change and rising sea levels, port infrastructure and shipping routes are becoming more and more susceptible to catastrophic weather events \cite{becker2013note}. Furthermore, operational risks such as equipment failures, delays in cargo handling, and human mistake in logistics planning can have a major impact on the efficiency of supply chains. Considering that over 80\% of worldwide trade is carried out by ships \cite{unctad}, any interruption to the movement of products around the world would be catastrophic. 

The suppliers' perspective is a crucial aspect of the marine supply chain, particularly with regard to the tools and equipment they provide that are essential to numerous nautical systems. This aspect of supply chain security is vital and warrants detailed examination to fully understand the scope of maritime supply chain security. The maritime sector includes a range of specialized technology, including cargo handling devices, propulsion systems, communication tools, and numerous sensors. A breach in one or more product supply chains might lead to many vulnerabilities in important maritime systems. Malicious actors attempting to install hardware or software intended to interfere with operations and obtain unauthorized access may take advantage of these vulnerabilities \cite{liu2023maritime}.  On the other hand, it may lead to the replacement of original parts with fake or inferior ones, which would raise the danger of equipment malfunctions and security issues \cite{yang2011risk}. Moreover, the introduction of malicious code into software or firmware during the manufacturing process may result in the creation of a backdoor that allows for unauthorized control \cite{wendler2020port}.

\subsection{Risk Assessment Method}
Risk assessment techniques offer a proactive approach to identifying and minimizing potential dangers, assist in preventing accidents, reduce potential financial losses, and comply with regulatory requirements. Unlike \cite{huang2023review}, this study does not attempt to comprehensively review every risk assessment technique used in the maritime context. Rather, we concentrate on a range of typical approaches and group them into five main categories: (1) data-driven; (2) semi-quantitative; (3) qualitative; (4) quantitative; and (5) emergent methods. 

The goal of quantitative methods is to provide numerical estimates of risk probabilities and consequences. These methods can be further classified into two categories: (a) artificial intelligence and machine learning techniques, such as Long Short-Term Memory (LSTM) networks \cite{zhang2024navigational}, and (b) probabilistic and statistical approaches, such as Bayesian networks \cite{baksh2018marine}. These techniques are helpful for comparing various hazards or evaluating the efficacy of risk reduction strategies, but they typically call for a significant amount of data. Accurately estimating the financial, operational, and reputational implications of cyber incidents can be difficult for those quantitative approaches \cite{kalogeraki2018novel}. Qualitative methods can be broadly classified into two categories: (a) hazard identification - Hazard and Operability Analysis (HAZOP) \cite{zhan2009application}; and (b) system-based approaches - Systems Theoretic Process Analysis (STPA) \cite{sultana2019hazard}. The former focuses on identifying and characterizing potential hazards and risks without necessarily quantifying them. These techniques are especially helpful for complex systems where quantification may be difficult and in the early phases of risk assessment. In order to provide a more nuanced assessment than simply qualitative methods, semi-quantitative methods mix qualitative and quantitative approaches. They typically use scoring or ranking systems without the full complexity of quantitative methods. They include (a) human factor analysis, or Human Factors Analysis and Classification System (HFACS) \cite{chauvin2013human}, and (b) failure analysis, or Failure Mode, Effects, and Criticality Analysis (FMECA) \cite{fu2014use}. Large datasets are a major component of data-driven approaches, which use them to find patterns, trends, and connections pertaining to marine threats. Examples include (a) association rules-based data mining and analysis techniques \cite{chen2022risk} and (b) geographic information system (GIS)-based geospatial analysis \cite{wang2022gis}. Additionally, more recent methods are beginning to emerge, including knowledge graphs \cite{gan2023knowledge}. Every approach has its own advantages and works well for various aspects of maritime cyber.

Nonetheless, the lack of an internationally recognized standard for maritime cyber risk assessment techniques could be the source of any discrepancies. Recent efforts have aimed to establish a comprehensive framework for risk analysis in maritime systems. For instance, International Association of Classification (IACS) Unified Requirement E22 \cite{e22} provides a structured approach to assessing the potential impact of failures, which is crucial for prioritizing risk mitigation efforts.  IACS Recommendation 171 \cite{IACS171} offers guidance on conducting systematic risk assessments, considering factors such as probability and consequence of potential hazards. These IACS initiatives represent significant steps toward standardization, but the complexity of maritime systems and the diversity of operational contexts continue to pose challenges. Above all, there might not be enough expertise with cybersecurity and maritime experience to perform comprehensive risk assessments \cite{alcaide2020critical}.

\subsection{Regulatory Frameworks/Laws}
IACS Unified Requirements E26 \cite{e26} and E27 \cite{e27} are comprehensive cybersecurity regulatory standards specifically tailored for the maritime industry. E26 focuses on ship-level cyber resilience, while E27 addresses the cyber resilience of individual onboard systems and equipment. They cover the entire life cycle of a vessel, from design and construction to operation and maintenance, and address key areas such as network segmentation, access controls, malware protection, and incident response. They help ship owners, system integrators, and manufacturers to implement robust cybersecurity measures, reducing the risk of cyber incidents. The International Ship and Port Facility Security (ISPS) Code \cite{code2016international} establishes an international framework for detecting security threats and taking preventive measures against security incidents affecting ships or port facilities used in trade. It defines roles and responsibilities for governments, local administrations, ship and port industries at the national and international level. 

However, these regulations do have potential drawbacks. Their complexity can be challenging for smaller operators or those with limited cybersecurity expertise to fully understand and implement. Additionally, the rapid pace of technological change in cybersecurity may challenge the ability of these standards to keep up with new threats. These regulations can become outdated quickly in the face of evolving cyber threats and create vulnerabilities in maritime cybersecurity. There has been issues in lack of standards for new technologies \cite{stalmokaite2023revival}. This shows how regulatory frameworks can lag behind technological innovations. Besides, many regulations do not provide detailed technical standards and leave much open to interpretation. Most importantly, they do not fully account for the rapid cyber threats and leave ships vulnerable to new attack vectors. For example, the ISPS Code was primarily designed to enhance maritime security against terrorism threats. Many shipping companies focused on meeting the minimum requirements of the ISPS Code. Ships compliant with ISPS might still be vulnerable to cyber attacks, as the compliance requirements did not comprehensively cover the evolving cyber threat landscape.

\subsection{Human Factor}
Empirical research suggests that human factors account for 65.8\% to 80\% of maritime incidents, with human-related events and actions being identified as the key contributing reason \cite{shi2021structured}. 
This diverse field includes a number of important elements: 1) personal cognitive aspects, such as mental workload, situation awareness, and decision-making techniques; 2) physiological elements, including exhaustion, stress, and irregularities in the circadian rhythm; 3) Interpersonal factors, which include issues with coordination, poor communication, and team dynamics; 4) the interplay between humans and machines, especially in light of the growing automation and technological complexity of maritime operations; 5) Organizational elements, such as management procedures, safety culture, and production pressure \cite{di2016human}\cite{galierikova2019human}\cite{grech2008human}.

Task performance is directly influenced by cognitive and physiological aspects at the microlevel. For example, it has been demonstrated that fatigue, a well-known problem in the maritime industry, impairs cognitive abilities including attention, decision-making, and situational awareness, which may result in mistakes made during navigation or a delay in reacting to urgent situations \cite{hetherington2006safety}. Interpersonal dynamics and the human-machine interaction significantly influence group performance at the mesolevel. Poor interface design can cause mode confusion or automation surprises, and inadequate team interactions can contribute to mistake chains that result in incidents \cite{chauvin2011human}. Human factors can play a significant influence in shaping corporate safety culture and policy formulation at a macro level. On the other hand, a weak safety culture may prioritize short-term operational efficiency ahead of long-term safety considerations \cite{haavold2009safety}. A strong safety culture can promote proactive risk management techniques and near-miss reporting. As a result, the incorporation of organizational, structural, individual, and team perspectives has highlighted the critical role that human factors play in the field of maritime cybersecurity.

\section{Maritime Cyber Attacks}

Academic research may focus on in-depth studies, analyzing attack patterns, 
investigating novel threat vectors, and exploring theoretical vulnerabilities 
that may not yet have been exploited in real-world scenarios. In contrast, 
industry reports often detail actual incidents, offering specifics about attack 
methodologies, targets, and outcomes that academic research might not have access 
to. We review existing cyber attacks, presented in both academic research and in industry,  
on the various maritime systems. 
%


\subsection{Real-Life Maritime Incidents (Industry)}
\label{sec:threats a}
Various reports on maritime cyber issues in recent years indicate an increase in both the number and sophistication of cyber incidents targeting shipping. These individual attacks range from ransomware against port systems to global GPS spoofing affecting vessel navigation. While precise incident numbers may vary due to different reporting methods and definitions of cyber incidents, there is a general consensus that the maritime sector has experienced a rise in threats.
\subsubsection{Multi-dimensional Attack Taxonomy}
\begin{figure*}[h]
  \centering
  \includegraphics[width=18cm]{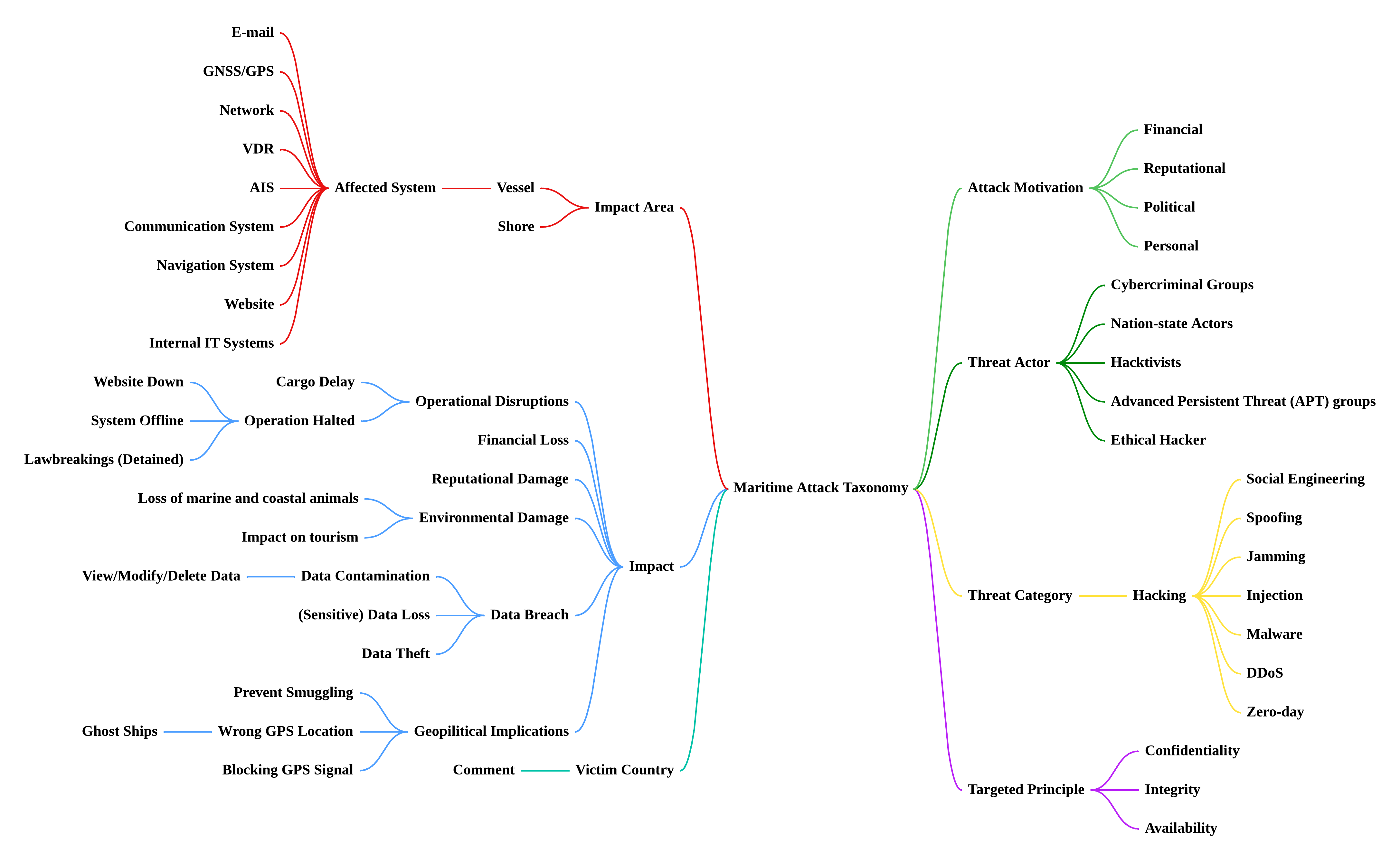}
  \caption{Multi-dimensional Maritime Cyber Attack Taxonomy}
  \label{fig:mindmap}
\end{figure*}

A well-defined taxonomy based on these dimensions (See Figure \ref{fig:mindmap}) can provide stakeholders in the maritime industry—shipbuilders, operators, and cybersecurity experts—with important new information about the types and extent of cyberthreats that affect the industry. A comprehensive analysis of all maritime incidents is outside the purview of this article, even though various studies have given an overview of noteworthy, publicly revealed cyber attack occurrences \cite{melandretrospective,orucclaims,schwarzstructured,senarak2023port}. The following significant factors have been taken into consideration while mapping the marine incidents into our multi-dimensional attack taxonomy:
\begin{itemize}

\item \textit{Threat Actor}: A threat actor is an entity, either external (e.g., cybercriminals, nation-states, hacktivists) or internal (e.g., disgruntled employees or contractors), that seeks to compromise the IT or OT security of a maritime system, network, or organization.
\item \textit{Impact Area}: The specific location or environment affected by a maritime incident, which can be categorized as vessel (e.g., onboard systems, cargo, and crew) and shore (e.g., port facilities and logistics). For affected systems: any shipboard IT or OT system, network, or equipment that has been compromised, damaged, or disrupted as a result of a maritime incident.
\item \textit{Attack Motivation}: The underlying reasons or incentives driving a threat actor to target and compromise a maritime system, network, or organisation. Common motivations include financial gain, espionage, political objectives, competitive advantage or the desire to disrupt operations.
\item \textit{Impact}: The negative effects or consequences of a cybersecurity incident on a maritime organization. The exact impact could result in an excessive amount of terms but it can be broadly categorized into operational (e.g., interruption of critical functions), political (e.g., territorial dispute), environmental (e.g., pollution), financial (e.g., ransom payments), reputational (e.g., damage to the brand), safety (e.g., crew), legal and regulatory (e.g., violation of laws). 
\item \textit{Targeted Principle}: The main intention of a threat actor's compromise of a network, organization, or marine system. The three core concepts of the aim are availability, integrity, and confidentiality. Unauthorized access to sensitive information is referred to as confidentiality. Unauthorized alteration or tampering with data or systems is referred to as integrity. Disruption or lack of access to essential resources is referred to as availability.
\item \textit{Victim Country}: The determination of jurisdiction may depend on whose territorial waters the ship was in at the time of the incident, or under which flag the affected ship is registered. For the purposes of reporting, establishing jurisdiction, and facilitating international collaboration in incident response and investigation, it may be necessary to identify the victim nation.
\item \textit{Threat Category}: A grouping of cybersecurity threats that share similar characteristics such as the type of attacker, the method of attack, or the target of the attack.
\end{itemize}


\subsubsection{Evaluation of Incidents by 2014 - 2023}
\begin{figure}[h]
  \centering
  \includegraphics[width=9cm]{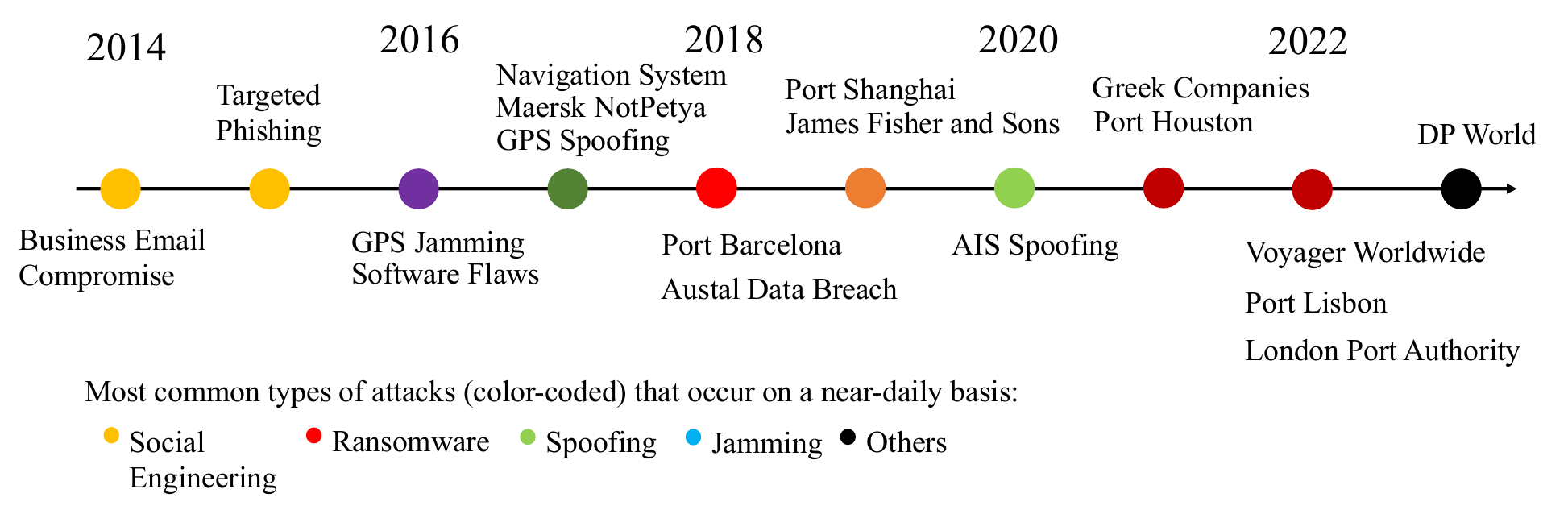}
  \caption{The timeline of the significant maritime cyber attack incidents with color representations}
  \label{fig:timeline}
\end{figure}

Cybercriminals have been targeting the maritime sector, which includes ports, ships, shipping firms, and maritime authorities, more and more in recent years. The maritime industry is becoming more and more susceptible to sophisticated cyberattacks, thus recent occurrences need to be closely examined. This section therefore concentrates on cyberattacks that have happened in the last ten years, specifically from 2014 to 2023. A thorough analysis has been conducted on the twenty chosen maritime events (see Table~\ref{tab:evaluate cyberattacks} for the evaluation and Figure~\ref{fig:timeline} for the timeline). Our goal is to contribute to a better understanding of the changing cyber threat landscape by exploring particular categories of contributing elements that may have been missed or insufficiently addressed in existing literature.

\textbf{Business Email Compromise (BEC) in Dubai} \cite{dubai}:
In 2014, Nautilus Minerals and Dubai-based marine solutions company Marine Assets Corporation (MAC) fell victim to a BEC attack, an attack that tricked Nautilus paying a \$10 million deposit intended for MAC into a fraudulent bank account.


\textbf{Targeted Phishing in Limassol} \cite{limassol}:
A Cyprus-based shipping company suffered a targeted phishing attack that resulted in a \$644,000 loss. The attacker carefully mimicked the legitimate hunker supplier's email communications and requested the funds to be sent to a different bank account. The fraud was discovered when the legitimate fuel company contacted the shipping company for the outstanding payment.


\textbf{GPS Jamming in South Korea} \cite{skjamming}:
In 2016, South Korea experienced a series of GPS jamming attacks that affected hundreds of ship, causing significant disruptions to navigation. The attacks, believed to have originated from hackers in North Korea, highlighted the vulnerability of GPS systems.

\textbf{Software Flaws in U.S. Ports} \cite{sqlus}:
Navis WebAccess, a web-based application essential for real-time operational data access in ports and logistics worldwide, was discovered to have a critical SQL injection vulnerability (CVE-2016-5817). Ethical hacker "bRpsd" publicly disclosed the vulnerability by releasing proof-of-concept (PoC) exploit code. Unfortunately, this disclosure occurred without prior notification to the vendor, potentially putting users at risk from malicious actors exploiting the vulnerability.


\textbf{Navigation System Attack} \cite{navigationspoof}: In February 2017, hackers seized control of a German-owned container vessel's navigation systems for 10 hours, rendering the captain unable to maneuver. The vessel's IT system was compromised, and onboard experts had to intervene to restore control. Industry sources suggest the attack was an attempt by pirates to divert the ship for boarding and ransom. 

\textbf{Maersk NotPetya Attack} \cite{maersk}: In 2017, the state-backed hacker group Sandworm, infamous for the NotPetya ransomware, launched a devastating cyberattack on Ukrainian businesses and organizations. It was facilitated by a widely used tax accounting software - ME Doc, and NotPetya spread rapidly, crippling computer systems not just in Ukraine, but also at global port terminals controlled by Maersk division. The attack infiltrated nearly 45,000 devices and 4,000 servers across 600 Maersk locations worldwide. This widespread disruption resulted in an estimated \$250-300 million loss in revenue for Maersk.

\textbf{GPS Spoofing in the Black Sea} \cite{gpsSpoofing}: In 2017, alarm bells rang out across the Black Sea when over 20 vessels reported their GPS positions as being inland at an airport, reported by U.S. Maritime Administration. This wasn't an isolated glitch – over 20 ships reporting the same false location, coupled with their positions bouncing back and forth between the airport and their true locations, strongly suggests a deliberate and large-scale GPS spoofing attack.



\textbf{Port of Barcelona Attack} \cite{barcelona}: In 2018, the Port of Barcelona in Spain experienced a significant cyber attack that affected its IT systems and caused disruptions to its operations. Due to system instability caused by the compromise, cargo handling efficiency between vessels and trailers significantly decreased, causing major delays in delivery. 

\textbf{Austal Data Breach} \cite{austal}: In mid-October 2018, Austal, Australia's shipbuilder and defence contractor, suffered a major data breach. Attackers purchased stolen Austal login credentials on the dark web and these credentials were used to gain access to Austal's Australian business data management systems. The attackers attempted to extort Austal by offering to return to the stolen data in exchange for a ransom payment.


\textbf{GNSS Spoofing in Port of Shanghai} \cite{shanghai}: In 2019, several incidents of GNSS (Global Navigation Satellite System) interference were reported in Chinese coastal areas, including the Port of Shanghai. This interference can disrupt or falsify GNSS signals, potentially affecting a vessel's navigation and communication equipment. It has been proposed that the Chinese government might be behind the GPS spoofing incidents, either as a security measure to conceal oil terminals and important government facilities or to evade surveillance of oil imports.

\textbf{James Fisher and Sons (JFS)} \cite{ukcompany}: UK-based marine services provider JFS disclosed an unauthorized intrusion into its computer systems. The cyberattack triggered a 5.7\% decline in the company's market shares.


\textbf{AIS Spoofing in Polish Waters} \cite{aispolish}: In November 2020, the Automatic Identification System (AIS) location data for the USS Roosevelt was spoofed. Threat actors, potentially linked to Russia, manipulated the AIS data to show the vessel near the Russian enclave of Kaliningrad, when it was actually elsewhere. The false AIS data could serve to create a narrative portraying Russia as the victim of encroaching naval activity.

\textbf{Death Kitty Ransomware} \cite{southafrica}: Transnet, a South Africa-based transport company, recently fell victim to a Death Kitty ransomware attack targeting its computer and NAVIS systems. This cyber incident has resulted in significant disruptions to Transnet's operations, with potential impacts lasting up to a week. The attack specifically targeted the company's port operations, leading to a complete seizure of port activities and a halt in the movement of cargo until the affected systems are fully restored.

\textbf{Attack on Greek Companies} \cite{greek}: In 2011, the companies impacted by the cyberattack leveraged the communication systems provided by Danaos Management Consultants, based in the coastal community of Piraeus in Greater Athens. This attack disrupted their ability to communicate with various stakeholders, including ships, suppliers, agents, charterers, and other key partners. 

\textbf{Attack on the Port of Houston} \cite{houston}: The Port of Houston in the USA encountered a cyber attack targeting its computer network. The attackers attempted to exploit a zero-day vulnerability, indicating that the flaw was not previously known to the software creator. The nature of the attack suggested the involvement of a nation-state actor, whose objective was to obtain sensitive government information and potentially disrupt or halt operations.


\textbf{Attack on Voyager Worldwide} \cite{vdrattack}: In 2022, Voyager Worldwide, a leading maritime technology company, fell victim to a cyber attack that resulted in the complete shutdown of its navigation services and solutions. This major incident affected over 1,000 shipping companies worldwide that rely on Voyager's technology and services.

\textbf{Ransomware Attack on the Port of Lisbon} \cite{lisbon}: The Port of Lisbon suffered a crippling ransomware attack and data breach on Christmas Day 2022. The LockBit group, known for their aggressive tactics, stole a trove of sensitive data including financial reports, cargo information, and ship logs. 
The Port of Lisbon has not publicly commented on the attack or whether they paid the ransom demands of \$1.5 million. 

\textbf{DDoS Attack on London Port Authority Websites} \cite{london}: The Port of London Authority (PLA) fell victim to a cyberattack which has knocked its website offline in May 2022. This incident disrupted the normal online operations and services provided by the PLA, impacting its ability to communicate important information, handle inquiries, and facilitate digital transactions.


\textbf{Ransomware Attack on the Port of Nagoya} \cite{japanransom}: In July 2023, the bustling port of Nagoya in Japan faced a crippling ransomware assault orchestrated by the Lockbit group. The cyberattack effectively paralyzed the port's computerized container handling system, halting the flow of incoming shipping containers for a duration of two days.

\textbf{DP World Australia Attack} \cite{dpworld}: In November 2023, DP World, a key player managing 40\% of Australia's maritime freight, announced a suspension of operations at its port terminals in Sydney, Melbourne, Brisbane, and Fremantle due to a cybersecurity breach. Responding to the cyber threat, the company took immediate action by isolating its computer systems from external networks, effectively shutting down operations. This resulted in a standstill, with approximately 30,000 shipping containers being left stranded.

\subsubsection{Lesson Learned Summary}
Two significant initiatives have emerged to document cybersecurity threats in the maritime sector. Researchers at NHL Stenden University in the Netherlands have compiled the Maritime Cyber Attack Database (MCAD) \cite{mcad}, a comprehensive archive cataloging over 160 discrete cyber attacks on the maritime transportation sector. Concurrently, Advanced Dataset of Maritime cyber Incidents ReleAsed for Literature (ADMIRAL) has developed a more extensive collection, centralizing data on over 470 cybersecurity incidents within the maritime sector. 
\begin{figure}[h]
  \centering
  
  \includegraphics[width=9cm]{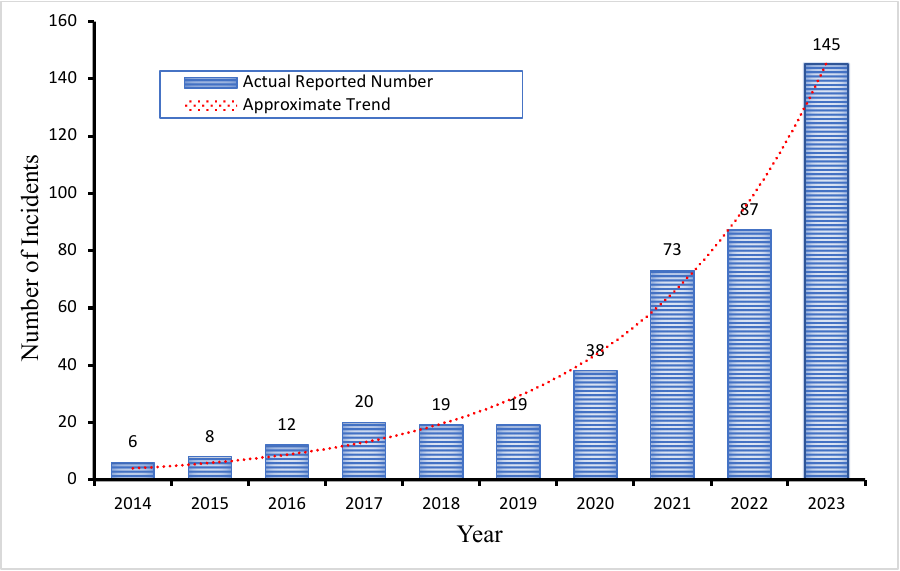}
  \caption{Maritime cyber incidents by year 2014 - 2023. Recreated by author based on data from MCAD (https://maritimecybersecurity.nl/) and ADMIRAL dataset (https://www.m-cert.fr/admiral/index.html). Each year's data is selected from the larger of the two sources.}
  
  \label{fig:marinetrend}
\end{figure}

\begin{table*}[h]\scriptsize

  \caption{Evaluation of Real-Life Maritime Cyber Attack Incidents}

  \label{tab:evaluate cyberattacks}
  \begin{tabular}{cp{3cm}p{0.9cm}p{0.9cm}p{0.9cm}cp{1.6cm}p{1.1cm}p{1.2cm}c}
    \toprule
    Year&Name&Victim Country&Impact Area&Affected System&Threat Category&Threat Actor&Attack Motivation&Impact&Targeted Principle\\
    \toprule
    2014 & Business Email Compromise (BEC) in Dubai & UAE & Shore &E-mail & Social Engineering & Cybercriminal Groups & Financial Gain & Financial Loss& Confidentiality\\
    \midrule
    2015 & Targeted Phishing in Limassol & Cyrus & Shore &E-mail & Phishing & Cybercriminal Groups & Financial Gain & Financial Loss& Confidentiality\\
    \midrule
    2016 & GPS Jamming in South Korea & South Korea & Vessel &GPS & Jamming & Nation-state Actors & Political Objectives & Blocking GPS signal & Availability\\
    \midrule
    2016 & Software Flaws in U.S. Ports & USA & Shore &Website & SQL injection attack & Ethical Hacker & Personal Entertain & View/Modify Operational Data& Integrity\\
    \midrule
    2017 & Navigation System Attack & Germany & Vessel &Navigation System & Hijacking & Cybercriminal Groups & Financial Gain & Unauthorized Control & Availability\\
    \midrule
    2017 & Maersk NotPetya Attack & Ukraine & Shore &Network & Malware & APT Groups & Political Objectives & Financial Loss & Confidentiality\\
    \midrule
    2017 & GPS Spoofing in the Black Sea & Russia & Vessel &GPS & GPS Spoofing & Nation-state Actors & Political Objectives & Wrong GPS Location & Integrity\\
    \midrule
    2018 & Port of Barcelona Attack & Spain & Shore &IT \makecell{Systems} & Ransomware & Unknown & N/A & Cargo \makecell{Delay} & Availability\\
    \midrule
    2018 & Austal Data Breach & Australia & Shore & Network & Ransomware & Cybercriminal Groups & Financial Gain & Data Theft& Confidentiality\\
    \midrule
    2019 & GNSS Spoofing in Port of Shanghai  & China & Shore &GNSS & GNSS Spoofing & Nation-state Actors & Reputation & Prevent Smuggling& Integrity\\
    \midrule
    2019 & James Fisher and Sons(JFS)& UK & Shore &Network & Ransomware & Cybercriminal Groups & Financial Gain & Reputational Damage& Availability\\
    \midrule
    2020 & AIS Spoofing in Polish Waters & USA & Vessel &AIS & AIS Spoofing & Nation-state Actors & Political Objectives & Ghost Ships & Integrity\\
    \midrule
    2021 & Death Kitty Ransomware & South Africa & Shore &Network & Ransomware & Cybercriminal Groups & Financial Gain & Operations Halted & Availability\\
    \midrule
    2021 & Attack on Greek Companies & Greece & Shore &Comm System & Ransomware & Cybercriminal Groups & Financial Gain & Data Loss&  Availability\\
    \midrule
    2021 & Attack on the Port of Houston & USA & Shore &Network & Zero-day & Nation-state Actors & Political Objectives & Sensitive Data Loss& Confidentiality\\
    \midrule
    2022 & Attack on Voyager Worldwide & Singapore & Shore &VDR & Hacking & Cybercriminal Groups & Financial Gain & Systems Offline& Availability\\
    \midrule
    2022 & Ransomware Attack on the Port of Lisbon & Portugal & Shore &Network & Ransomware & Cybercriminal Groups & Financial Gain & Financial Loss& Availability\\
    \midrule
    2022 & DDoS Attack on London Port Authority Websites & UK & Shore &Website & DDoS & Hacktivists & Political Objectives & Website Down& Availability\\
    \midrule
    2023 & Ransomware Attack on the Port of Nagoya & Japan & Shore &Network & Ransomware & Cybercriminal Groups & Sabotage & Operations Halted& Availability\\
    \midrule
    2023 & DP World Australia Attack & Australia & Shore &Network & Hacking & Cybercriminal Groups & Financial Gain & Data Theft& Availability\\

  \bottomrule

\end{tabular}

\end{table*} 
Both databases (which have not been updated since March 2024) reveal an alarming trend: the frequency of maritime cyber incidents have been steadily increasing from 2014 to 2023 (see Figure~\ref{fig:marinetrend}). According to the databases, cyber attacks in the maritime sector have targeted a wide array of systems and organizations. These include ship systems like navigation, port infrastructure, shipping companies, maritime authorities, government agencies, shipbuilders, and defense contractors. This diversity of targets highlights the broad attack surface within the maritime industry, spanning from individual vessels to large-scale port operations and regulatory bodies. However, the stagnation of these databases underscores the broader challenge of sustaining critical information resources. The need for consistent funding, dedicated personnel, and perhaps collaborative industry efforts to ensure the continual update should be highlighted.



As shown in Table \ref{tab:evaluate cyberattacks}, maritime cyber incidents have affected numerous countries and regions worldwide, including the United States, Europe (e.g., UK, Germany), Asian-Pacific (e.g., Singapore, Japan, Australia), and Middle East (e.g., UAE). This global distribution of attacks underscores that maritime cybersecurity is an international concern, with no country or region immune to these threats. The cyber attacks documented span a range of categories, including abuse and theft of data, DoS attacks, GPS spoofing and jamming, malware infections (particularly ransomware), social engineering and phishing attempts, targeted attacks, and exploitation of system vulnerabilities. This attack range shows how diverse the cyberthreats that the maritime industry faces are. These attacks are carried out by a similarly varied array of players, including nation-states, hacktivists, cybercriminal organizations, and lone hackers. The threat actors' identities remained a mystery in numerous instances. This diversity of perpetrators suggests that maritime cyberattacks are motivated by a variety of factors.

\subsection{Cyber Attack Demonstrations (Academic)}
\label{sec:threats b}
Academic cyber attack demonstrations typically take place in simulated environments or self-developed software that replicate maritime systems and networks. There is a portion of academic research focuses on theoretically feasible attacks by modeling maritime systems to identify potential attack vectors, even if they have not been practically implemented in a full-scale simulation. These demonstrations often focus on specific components of maritime systems 

\textbf{AIS Threats Overview}
Balduzzi \textit{et al.} \cite{balduzzi2014security} conducted a comprehensive security evaluation of the AIS and categorized these threats into three main types: i) spoofing, ii) hijacking, and iii) availability disruption attacks. Kessler \textit{et al.} \cite{kessler2018taxonomy} presented a comprehensive analysis of the potential cyber threats to the AIS by drawing upon previous research \cite{gauthier2018addressing}\cite{purton2010identification}\cite{strohmeier2014security} that examined the vulnerabilities of Automatic Dependent Surveillance-Broadcast (ADS-B), a similar system used in the aviation industry to provide situational awareness for aircraft. Amro and Gkioulos \cite{amro2022click} discussed a novel cyber attack that utilizes the AIS as a covert channel for sending command and control messages and delivering malware to maritime systems.

\textbf{VDR Threats Overview}
Harish \textit{et al.} \cite{harish2022investigating} launched a multi-pronged attack on a commercially available VDR, employing a USB Rubber Ducky for physical infiltration, the Metasploit framework for system exploit, a ransomware simulator to demonstrate data encryption risks, and Nmap to identify exploitable network vulnerabilities. Hopcraft \textit{et al.} \cite{hopcraft2023raising} highlighted cybersecurity deficiencies in current VDR systems and standards, and presented they are susceptible to ransomware attacks, malware infections, denial-of-service (DoS) attacks, insider threats and data breaches. S{\"o}ner \textit{et al.} \cite{soner2023cybersecurity} used failure modes and effects analysis (FMEA) to assess cyber vulnerabilities and potential attacks against VDR. Data acquisition unit (DAU) is found to be the most vulnerable VDR component, followed by remote access solutions and the bridge control panel. False data injection, command injection, and viruses are ranked as the top three threats in the FMEA risk assessment.

\textbf{RADAR Threats Overview}
Junior \textit{et al.} \cite{leite2021triggering} demonstrated how the attacker could send the command to the ship's radar via an electronic attack (EA) that generates false radar echoes in a specific pattern. Wolsing \textit{et al.} \cite{wolsing2022network} presented a taxonomy of possible network cyber attacks on marine RADAR and a simulation environment to implement these attacks. The taxonomy covers denial of service attacks, attacks involving basic image transformations like scaling, rotation and translation, and targeted object manipulations like addition, removal and relocation of radar echoes. Longo \textit{et al.} \cite{longo2023attacking} demonstrated that it is feasible to develop malware that can autonomously and stealthily target radar systems on ships by exploiting vulnerabilities in standard protocols like NMEA and ASTERIX that lack authentication and encryption. Longo \textit{et al.} \cite{longo2023electronic} discussed a novel threat model that combines cyber attacks with cyber false flags to target maritime RADAR systems. The proposed cyber attacks manipulate RADAR data to simulate the effects of electronic countermeasures (ECM) like barrage jamming, spot jamming, digital radio frequency memory, and blip enhancement. By making the cyber attacks appear as ECM originating from another ship, it creates a false flag that misleads attribution of the attack. These false flag attacks aim to disrupt radar operations, divert blame, hide the cyber nature of the attack, and project the attacker's offensive capabilities. 

\textbf{VSAT Threats Overview}
Santamarta \cite{santamarta2014satcom} discussed security vulnerabilities found in various widely deployed satellite communications (SATCOM) terminals and described several real-world attack scenarios showing how these flaws could be exploited to leak sensitive military data, interfere with distress communications, spoof navigation data, disable safety systems, and more. Based on the two scenarios presented, VSAT systems in the maritime context face threats of data manipulation, spoofing, and communication disruption. Pavur \textit{et al.} \cite{pavur2020tale} revealed serious security flaws in how maritime VSAT networks are currently implemented. The insecure VSAT links could allow attackers to modify important data like ECDIS navigational charts and AIS vessel tracking information, impacting ship safety. Under certain conditions, an attacker could hijack active TCP sessions and modify data in real-time.

\textbf{GNSS Threats Overview}
Grant \textit{et al.} \cite{grant2009gps} conducted practical experiments to validate the threat of GPS jamming on maritime navigation. Bhatti and Humphreys \cite{bhatti2017hostile} analyzed and demonstrated the ability of an attacker to control a maritime surface vessel by broadcasting counterfeit civil GPS signals. Medina \textit{et al.} \cite{medina2019gnss} carried out a dedicated jamming measurement campaign on the Baltic Sea in cooperation with the German Federal Network Agency. 

\textbf{ECDIS Threats Overview}
All five papers written by Svilicic \textit{et al.} \cite{svilicic2019raising, svilicic2019assessing, svilicic2019maritime, svilicic2020paperless, svilicic2019shipboard} focused on the cyber security aspects of the ECDIS and used similar vulnerability scanning methods. However, each paper had a unique focus. Svilicic \textit{et al.} \cite{svilicic2019raising} presented an estimation of the cyber security vulnerabilities of ECDIS that arise from weaknesses related to the underlying operating system. Svilicic \textit{et al.} \cite{svilicic2019assessing} introduced a framework for assessing cyber risks of the ship's critical systems and assets. Svilicic \textit{et al.} \cite{svilicic2019maritime} identified the main threats to the ECDIS system through a survey of the ship's crew and computational vulnerability scanning using Nessus Professional software. Svilicic \textit{et al.} \cite{svilicic2020paperless} highlighted that interconnecting multiple ECDIS workstations with the same vulnerabilities provides an ideal environment for malware to spread. Svilicic \textit{et al.} \cite{svilicic2019shipboard} emphasized on the cyber security weaknesses of a ECDIS arising from the software's third-party components.
Dyryavyy \cite{ncc} discussed the cyber security risks and weaknesses within ECDIS and the research revealed that a malicious actor could potentially browse, download, modify, or erase any file on the computer running the ECDIS software \cite{hayes2016maritime}.

\textbf{GMDSS Threats Overview}
While specific GMDSS cyber attack examples are limited, the concepts of jamming, spoofing and DoS 
 \cite{akpan2022cybersecurity}\cite{tam2018maritime} discussed could all theoretically be used to disrupt GMDSS functionality as ships become more networked and vulnerable to cyber threats. The authors \cite{tam2018maritime} also discussed the risk of malware infections in ship systems. While not specifically mentioning GMDSS, such an attack could potentially affect GMDSS equipment if it's connected to infected systems.

\section{Security Solutions}

\subsection{Integrated or Holistic}
\label{sec:mitigation a}
Rather than focusing on individual components, integrated solutions aim to create a comprehensive framework that protects and optimizes the vessel as a whole. For example, an integrated solution might include a centralized security operations center that monitors and manages threats across all vessel systems, from navigation and communication to cargo management and environmental controls. Table \ref{tab:integrated} highlights the main holistic security solutions now in use, with an emphasis on contrasting their evaluation techniques and validation datasets.
\subsubsection{Machine/Deep Learning-Based Defence}
The demonstrated performance and adaptability of some advanced deep learning techniques suggest significant potential for their practical application in safeguarding critical maritime infrastructure against the ever-evolving landscape of cyber threats. In \cite{kumar2021dltif}, it was revealed that the DLTIF framework exhibited superior performance in automatically extracting and identifying cyber threat patterns compared to traditional machine learning approaches across multiple evaluation metrics. Dual Stack Machine Learning (S2ML) framework is proposed in \cite{ali2022securing} which takes a novel approach by leveraging entropy-based features extracted from network traffic, a method that proves particularly effective in identifying anomalies indicative of DDoS attacks. Similarly, the authors of \cite{gyamfi2022adaptive} leveraged the capabilities of an adaptive incremental passive-aggressive machine learning (AI-PAML) which is designed to continuously update its learning model as new network attacks are detected.

While DLTIF, S2ML and AI-PAML focus on identifying or detecting threats, the Multi-Agent Reinforcement Learning (MARL) \cite{wilson2024multi} approach takes the next step by exploring how AI can be used to actively defend against these threats. The MARL solution establishes a simulation environment for training autonomous cyber defence agents for maritime operational technology (OT) to enhance the resilience of maritime control systems.
\subsubsection{Cryptographic Schemes}
The distributed nature of maritime systems, with ships, ports, and other infrastructure being geographically dispersed, renders centralized security solutions impractical. Recent research has proposed several cryptographic schemes that provide security in a distributed manner, employing techniques such as symmetric encryption, attribute-based encryption, erasure coding, identity-based encryption, and digital signatures to enable each entity to safeguard its own data and communications. A lightweight authentication protocol \cite{chaudhry2021lightweight} using symmetric cryptography primitives like XOR and hash operations can provide mutual authentication and shared key establishment while ensuring vessel privacy. Similarly, an attribute-based secure data aggregation scheme \cite{wang2021attribute} employs constant attributes of maritime terminals for authentication and onboard sensors for data encryption, using zero-knowledge proofs for member certification. Addressing a critical aspect of maritime data security overlooked by the previous schemes, it was proposed that a flexible integrity checking and recovery mechanism \cite{liu2021flexible} is particularly valuable in maritime environments where harsh conditions frequently lead to data corruption. Other prevention methods include an identity-based authentication mechanism \cite{gupta2021identity} that integrates both cryptographic techniques and distributed ledger technologies to ensure not only the confidentiality and integrity of data but also enhance transparency and non-repudiation in maritime communications. 

The prevalence of legacy equipment in maritime communication channels, often predating modern security protocols and persisting due to extended lifecycles, presents significant cybersecurity vulnerabilities in today's interconnected maritime environment. In order to offer a pragmatic strategy for a gradual, cost-effective transition towards more secure maritime communication paradigms, the authors of \cite{hemminghaus2021sigmar} proposed the SIGMAR framework, which extends the IEC 61162-450 protocol and integrates the Elliptic Curve Digital Signature Algorithm (ECDSA) to provide robust authentication and integrity verification with minimal latency and communication overhead.

\subsubsection{Blockchain-Based Defence}
A secure communication framework can be harnessed by introducing a private blockchain network integrated with a terrestrial fusion center for authentication purposes \cite{rahimi2020secure}. The approach is designed to provide a resilient and trustworthy communication infrastructure for maritime systems, with particular emphasis on unmanned aerial vehicle (UAV)-assisted sensing applications in maritime environments. In \cite{zhang2022blockchain}, a blockchain-based authentication mechanism is utilized to store and validate the identities of vessels and IoT devices. The solution implements a proof-of-authority consensus mechanism and incorporates a threshold-based approach for detecting malicious nodes. Further, blockchain can be used to secure positioning data in maritime environments. A blockchain-based privacy preserving position data sharing approach is leveraged in \cite{gai2022blockchain} to address the critical privacy challenges in maritime IoT systems.

By harnessing the immutability and distributed nature of blockchain technology, these kinds of techniques aim to mitigate security vulnerabilities inherent in traditional centralized systems, while accommodating the unique challenges posed by maritime IoT ecosystems.
\subsubsection{Network Level-Based Defence}
Maritime vessels are equipped with systems of varying criticality levels \cite{rajaram2022guidelines}, with navigation and propulsion systems being mission-critical, in contrast to the non-essential nature of crew entertainment systems. A zone-based network segmentation approach, dividing ships' network into isolated zones like the Global Ship Zone, Ship Control Zone, and Ship System Zone, is proposed in \cite{furumoto2020toward} that help contain potential threats and prevent them from spreading to other critical areas. Further, a cyber attack path discovery method \cite{polatidis2018cyber} for maritime risk management can be used to identify potential vulnerabilities in network configurations. 

In \cite{he2021dns}, it was suggested that creating and maintaining whitelists of authorized devices and their permissions could provide a strong, regular security check for the local network environment of maritime IoT devices against DNS rebinding attacks.
\subsubsection{Miscellaneous / Hybrid}
Security solutions for maritime threats do not fit neatly into the sole categories of machine/deep learning, blockchain, network level or cryptographic approaches but rather integrate various methods to achieve their goals. A rule-based risk model \cite{fossier2014risk} utilizing geospatial analysis, anomaly detection, and statistical methods is introduced to generate risk maps and real-time alerts, aimed at detecting abnormal behaviors indicative of maritime attacks. 
The ISOLA project \cite{laso2022isola} integrates a broad spectrum of technologies and methodologies, including sensor networks, data fusion, semantic reasoning, and visual analytics, while incorporating cyber vulnerability assessment tools and decision support systems to aid security personnel in identifying and responding to potential threats in the cruise ship industry.
Chen and Wu \cite{chen2022automatic} combined various technologies such as AIS, deep learning and cryptographic schemes to address multiple issues in maritime such as data integrity, authentication and network security.
\begin{table*}[]
\centering
\footnotesize
\caption{The Evaluation of Existing Integrated/Holistic Countermeasures Against Most Common Threats Based on Maturity, Dataset, and Evaluation Method}
\label{tab:integrated}
\begin{tabular}{lllllllll}
\toprule
\multirow{2}{*}{Approach} &
  \multicolumn{1}{c}{\multirow{2}{*}{Asset}} &
  \multicolumn{1}{c}{\multirow{2}{*}{Maturity}} &
  \multicolumn{2}{c|}{Dataset} &
  \multicolumn{3}{c}{Evaluation Method}  &
  \multicolumn{1}{c}{\multirow{2}{*}{*Performance}} \\ \cline{4-8}
 &
  \multicolumn{1}{c}{} &
  \multicolumn{1}{c}{} &
  \multicolumn{1}{c}{Pre-obtained} &
  \multicolumn{1}{c|}{Generated} &
  \multicolumn{1}{c}{Proof-of-Concept} &
  \multicolumn{1}{c}{Simulation} &
  \multicolumn{1}{c}{Others} \\ \toprule
\multirow{4}{*}{ML/DL} & \cite{kumar2021dltif}
  & Experimental
   & \multicolumn{1}{c}{\checkmark}
   &
   & \multicolumn{1}{c}{\checkmark}
   &
   &
   & \multicolumn{1}{c}{\righthalfcircle}
   \\ \cline{2-9}
 &
   \cite{ali2022securing}
  & Experimental
   & \multicolumn{1}{c}{\checkmark}
   &
   & \multicolumn{1}{c}{\checkmark}
   & \multicolumn{1}{c}{\checkmark}
   &
   &\multicolumn{1}{c}{\LEFTcircle}
   \\ \cline{2-9}
 &
   \cite{gyamfi2022adaptive}
  & Experimental
   & \multicolumn{1}{c}{\checkmark}
   & \multicolumn{1}{c}{\checkmark}
   & \multicolumn{1}{c}{\checkmark}
   & \multicolumn{1}{c}{\checkmark}
   &
   &\multicolumn{1}{c}{\righthalfcircle}
   \\ \cline{2-9}
 &
   \cite{wilson2024multi}
  & Experimental
   &
   &
   &
   & \multicolumn{1}{c}{\checkmark}
   &
   &\multicolumn{1}{c}{\righthalfcircle}
   \\ \hline
\multirow{5}{*}{Crypto} & \cite{chaudhry2021lightweight}
  & \multicolumn{1}{c}{Theoretical}
   &
   &
   &
   &
   & \multicolumn{1}{c}{\checkmark}
   &\multicolumn{1}{c}{$\mathord{?}$}
   \\ \cline{2-9}
 &
   \cite{wang2021attribute}
  & \multicolumn{1}{c}{Theoretical}
   &
   &
   & \multicolumn{1}{c}{\checkmark}
   &
   &
   &\multicolumn{1}{c}{$\mathord{?}$}
   \\ \cline{2-9}
 &\cite{liu2021flexible}
  & Experimental
   &
   &
   &
   & \multicolumn{1}{c}{\checkmark}
   &
   &\multicolumn{1}{c}{\righthalfcircle}
   \\ \cline{2-9}
 &\cite{gupta2021identity}
  & \multicolumn{1}{c}{Theoretical}
   &
   &
   & \multicolumn{1}{c}{\checkmark}
   &
   &
   &\multicolumn{1}{c}{$\mathord{?}$}
   \\ \cline{2-9}
 &
   \cite{hemminghaus2021sigmar}
  & Experimental
   &
   &
   & \multicolumn{1}{c}{\checkmark}
   & \multicolumn{1}{c}{\checkmark}
   &
   &\multicolumn{1}{c}{\righthalfcircle}
   \\ \hline
\multirow{3}{*}{Blockchain} & \cite{rahimi2020secure}
  &Experimental
   &
   &
   &
   &\multicolumn{1}{c}{\checkmark}
   &
   &\multicolumn{1}{c}{\righthalfcircle}
   \\ \cline{2-9}
 &
   \cite{zhang2022blockchain}
  &Experimental
   &
   &
   &
   &\multicolumn{1}{c}{\checkmark}
   &
   &\multicolumn{1}{c}{\righthalfcircle}
   \\ \cline{2-9}
 &
   \cite{gai2022blockchain}
  &Experimental
   &
   &
   &
   &\multicolumn{1}{c}{\checkmark}
   &
   &\multicolumn{1}{c}{\LEFTcircle}
   \\ \hline
\multirow{3}{*}{Network} &\cite{furumoto2020toward}
  &\multicolumn{1}{c}{Theoretical}
   &
   &
   &
   &
   & \multicolumn{1}{c}{\checkmark}
   &\multicolumn{1}{c}{$\mathord{?}$}
   \\ \cline{2-9}
 &\cite{polatidis2018cyber}
  &Experimental
   & \multicolumn{1}{c}{\checkmark}
   & \multicolumn{1}{c}{\checkmark}
   & \multicolumn{1}{c}{\checkmark}
   &
   &
   &\multicolumn{1}{c}{\righthalfcircle}
   \\ \cline{2-9}
 &\cite{he2021dns}
  &Experimental
   &
   &
   &
   & \multicolumn{1}{c}{\checkmark}
   &
   &\multicolumn{1}{c}{\righthalfcircle}
   \\ \hline
\multirow{3}{*}{Hybrid} &\cite{fossier2014risk}
  &Experimental
   &\multicolumn{1}{c}{\checkmark}
   &\multicolumn{1}{c}{\checkmark}
   &
   &\multicolumn{1}{c}{\checkmark}
   &
   &\multicolumn{1}{c}{\righthalfcircle}
   \\ \cline{2-9}
 &
   \cite{laso2022isola}
  &\multicolumn{1}{c}{Theoretical}
   &
   &
   &
   &
   &\multicolumn{1}{c}{\checkmark}
   &\multicolumn{1}{c}{$\mathord{?}$}  
   \\ \cline{2-9}
 &
   \cite{chen2022automatic}
  &Experimental
   &
   &
   &\multicolumn{1}{c}{\checkmark}
   &\multicolumn{1}{c}{\checkmark}
   &
   &\multicolumn{1}{c}{\LEFTcircle}
   \\ \bottomrule
\multicolumn{8}{l}{Pre-obtained = Dataset already exists, and is not created by the researcher themselves}\\
\multicolumn{8}{l}{Generated = Dataset created specifically for the research at hand, through tools and methods they control}\\
\multicolumn{8}{l}{Others = Its efficiency was evaluated through theoretical assumptions or comparisons without experimental validation} \\
\multicolumn{8}{l}{\CIRCLE = High \LEFTcircle = Medium \Circle = Low \righthalfcircle = Medium to High \lefthalfcircle = Low to Medium $\mathord{?}$ = Unsure as the author did not carry out validations}\\
\multicolumn{8}{l}{*Performance is assessed through distinct metrics, which includes compatibility, implementation costs, robustness and so on}\\
\bottomrule
\end{tabular}
\end{table*}

\subsection{Component Specific}
\label{sec:mitigation b}
Component-specific solutions refer to targeted countermeasures or protective measures designed to address vulnerabilities or prevent threats associated with particular vessel components. These solutions are tailored to the unique characteristics and vulnerabilities of each component, taking into account its function, exposure to threats, and critical role in the vessel's overall operation. We have a summarized list here with additional comparative study of different approaches in Table \ref{tab:component}.

\subsubsection{AIS Countermeasures} Kowalska and Peel \cite{kowalska2012maritime} presented an approach for detecting anomalous vessel behaviour using Gaussian Process (GP) models trained on AIS data. The GP anomaly detection method focuses on identifying suspicious vessel behaviors rather than securing the communication channel or data itself. Similarly, the work in \cite{ray2015deais} also focuses on identifying potential anomalies by processing incoming AIS data. Balduzzi \textit{et al.} \cite{balduzzi2014security} recommended applying anomaly detection techniques to the AIS data to identify suspicious activities, such as unexpected changes in a vessel's route, which could indicate spoofing or hijacking attempts. The author \cite{balduzzi2014security} also suggested to implement a Public Key Infrastructure (PKI) using X.509 certificates to enable authentication and integrity checks for AIS messages. Influenced by security solutions from IEEE 1609 standards for vehicle communications, Hall \textit{et al.} \cite{hall2015ieee} proposed a redesigned security protocol to address multiple security vulnerabilities in the current AIS vessel tracking system. To identify vessels that deliberately turn off their AIS transponders, Bernab{\'e} \textit{et al.} \cite{bernabe2023detecting} used a self-supervised deep learning approach to analyze patterns in AIS transmissions. By comparing these predictions to actual real-time observations, it can classify vessel trajectories as normal or potential suspicious. Iphar \textit{et al.} \cite{iphar2015detection} outlined a conceptual methodology for using various data quality dimensions to assess the integrity and trustworthiness of information within AIS messages. Su \textit{et al.} \cite{su2017privacy} proposed authentication and privacy-preserving enhancements to the AIS protocol to address the security vulnerabilities that currently allow identity spoofing and enable tracking of vessels through their AIS broadcasts. They first introduced Digital Signature based Identity Authentication Scheme (DSIAS) to prevent tampering or forging of vessel identities and then proposed Mix-zone based Trajectory Privacy Protection Scheme (MTPPS) to provide anonymity of vessel trajectories. They further built a Blind-signature extension to the MTTPS, which provides an additional layer of privacy. Kontopoulos \textit{et al.} \cite{kontopoulos2018countering} advocated an architectural solution for detecting malicious tampering of live AIS data streams. Goudossis and Katsikas \cite{goudossis2019towards} explored how a Maritime Certificate-less Identity-Based Public Key Cryptography (mIBC) infrastructure may enhance the security properties of AIS. While the previous paper introduced the concept of using Identity-Based cryptography to secure AIS, Goudossis and Katsikas \cite{goudosis2020secure} further delved into the implementation details, introduced a new application for seamless integration with the existing AIS infrastructure, and provided operational overhead estimates. Nguyen \cite{nguyen2020hardening} evaluated the AIS architecture and explored using lightweight cryptographic algorithms to design an optimal authentication system. Sciancalepore \textit{et al.} \cite{sciancalepore2021auth} focused on the design, security properties, implementation and performance evaluation of the proposed Auth-AIS protocol to secure vessel AIS broadcast messages in a practical manner. Kelly \cite{kelly2022novel} used Received Signal Strength Indicator (RSSI) based detection method to enhance AIS security by providing a means to identify and investigate vessels attempting to exploit the low power mode to evade tracking, which could indicate involvement in illicit activities. Deng \textit{et al.} \cite{deng2023lightweight} proposed a novel lightweight Transformer-based network called GLFormer for specific emitter identification (SEI) to provide an extra security layer for AIS terminal emitters.


\subsubsection{GNSS Countermeasures}
Grant \textit{et al.} \cite{grant2009gps} tested the feasibility of eLoran as a backup system to provide PNT during GPS jamming incidents. Bhatti and Humphreys \cite{bhatti2017hostile} proposed several countermeasures to mitigate the risk of GPS spoofing attacks on maritime vessels. 
In contrast, Drumhiller \textit{et al.} \cite{drumhiller2017issues} focused on primary defenses against GPS jamming. Medina \textit{et al.} \cite{medina2019gnss} proposed jamming countermeasures like robust signal processing, adaptive antenna arrays, or multi-sensor fusion. A new technique for detecting GNSS spoofing attacks using signals from the Iridium satellite constellation was proposed by Oligeri \textit{et al.} \cite{oligeri2020gnss}. Spravil \textit{et al.} \cite{spravil2023detecting} focused on detecting GPS spoofing attacks in the maritime domain using a novel software-based framework called MANA (MAritime Nmea-based Anomaly detection). Boudehenn \textit{et al.} \cite{boudehenn2021navigation} proposed a strategy to enhance the detection of navigation spoofing attacks and assess possible physical impacts.

\subsubsection{ECDIS Countermeasures}
Based on the information provided by Svilicic \textit{et al.} and Dyryavyy, some mitigation measures can be implemented to enhance the security of ECDIS systems. While the papers \cite{ncc}\cite{svilicic2019raising, svilicic2019assessing, svilicic2019maritime, svilicic2020paperless, svilicic2019shipboard} offer valuable insights into ECDIS threats and propose mitigation measures, the authors did not include experiments specifically designed to test the effectiveness of those mitigation.
These theoretical measures can be categorized into technical, procedural, and organizational aspects.

\subsubsection{RADAR Countermeasures}
Various security solutions and mitigation measures have been developed to protect RADAR systems, both in general and for specific applications, although they are not exclusive to the maritime domain. These include deep learning-based anomaly detection techniques \cite{cohen2022radarnomaly}\cite{de2022anomaly}, hash-based integrity checks and encryption designed for the ASTERIX protocol \cite{casanovas2015vulnerability}, and algorithmic approaches \cite{yang2018novel}. Despite the existence of these security solutions, it can be challenging to directly apply them to the maritime scenarios \cite{wolsing2022network}. The unique characteristics and requirements of marine RADAR systems should necessitate tailored approaches to ensure effective security measures. Junior \textit{et al.} \cite{leite2021triggering} briefly mentioned that as future work, they may develop tools to verify the integrity of the software used in naval RADAR systems, in order to detect any malicious code or malware that may have been pre-installed in the RADAR. Longo \textit{et al.} \cite{longo2023attacking} proposed a detection system as a countermeasure to the marine RADAR attacks. It is designed as a policy enforcement system where the policies dictate how the RADAR should operate according to industry standards, regulations, and manufacturer specifications, in conjunction with onboard configurations. The experimental results demonstrate that the detection system proposed in this study consumes fewer resources while detecting these attacks with high accuracy.

\subsubsection{VDR Countermeasures}
S{\"o}ner \textit{et al.} \cite{soner2023cybersecurity} also recommended several preventive and control measures to improve the cybersecurity of VDR based on their FMEA risk assessment findings. They emphasized that while cyberattacks cannot be entirely prevented due to the nature of the cyber world, their effects can be mitigated by conducting cyber risk assessments and implementing effective control measures to safeguard VDR from current and emerging cybersecurity threats. The authors \cite{hopcraft2023raising} also proposed several amendments to the standards to improve VDR security.

\subsubsection{VSAT Countermeasures}
The author \cite{pavur2020tale} acknowledged that some of these solutions, such as end-to-end encryption, may have performance implications due to the high latency of satellite communications. However, they emphasized the need for the maritime industry to prioritize security and invest in the development of practical, satellite-optimized security measures to protect ships, crew, and cargo from potential cyber threats. Wu \textit{et al.} \cite{wu2018approach} proposed a new lightweight authentication scheme called lite-CA (Lite Certification Authority) to reduce the amount of interaction required for authentication within the VSAT network architecture. To achieve real-time data encryption, they further proposed using a lightweight encryption algorithm called HW-F (high weight function) to replace traditional public key and symmetric encryption systems.

\subsubsection{GMDSS Countermeasures}
Korcz \cite{korcz2023key} discussed the ongoing modernization of GMDSS which includes introduction of new satellite providers beyond Inmarsat (e.g. Iridium and Beidou), development of the VHF Data Exchange System (VDES) for improved data communication and implementation of digital broadcasting of maritime safety information. These measures could provide more secure and efficient data exchange and enhance GMDSS communications. Os{\'e}s and Juncadella \cite{martinez2021global} argued that it is conceptually and technologically feasible to create a global VTS system as a counterpart to GMDSS.



\begin{table*}[]
\footnotesize
\centering
\caption{The Evaluation of Existing Component Specific Countermeasures Against Most Common Threats Based on Maturity, Evaluation Method, and Performance}
\label{tab:component}
\begin{tabular}{llllllll}
\toprule
\multirow{2}{*}{Category} &
  \multirow{2}{*}{Attacks Mitigated} &
  \multirow{2}{*}{Approach}&
  \multicolumn{1}{c}{\multirow{2}{*}{Maturity}} &
  \multicolumn{3}{c}{Evaluation Method} &
  \multirow{2}{*}{*Performance} \\ \cline{5-7}
 & 
   &
   &
  \multicolumn{1}{c}{}&
  \multicolumn{1}{c}{PoC} &
  \multicolumn{1}{c}{Simulation} &
  \multicolumn{1}{c}{Others} &
   \\ \toprule
\multirow{13}{*}{AIS} &
  \multirow{13}{*}{\begin{tabular}[c]{@{}l@{}}AIS Spoofing\\ Drug Smuggling\\ People Trafficking\\ Trajectory Tracking\\ Illegal Fishing\\ AIS Message Injection\\ AIS Message Flooding\\ AIS Data Tampering\\ Stream Poisoning \\ Unusual Vessel Patterns\\ Availability Disruption\\ AIS Hijacking\end{tabular}} &
  Detection model  \cite{kowalska2012maritime}&
    \multicolumn{1}{c}{Experimental}&
  \multicolumn{1}{c}{} &
  \multicolumn{1}{c}{\checkmark} &
  \multicolumn{1}{c}{} & 
  \multicolumn{1}{c}{\LEFTcircle}
   \\ \cline{3-8} 
 &
   &
  Detection technique \cite{ray2015deais}&
  \multicolumn{1}{c}{Theoretical}&
   &
   &
   \multicolumn{1}{c}{\checkmark}& 
   \multicolumn{1}{c}{$\mathord{?}$}
   \\ \cline{3-8} 
 &
   &
  PKI using X.609 \cite{balduzzi2014security} &
   \multicolumn{1}{c}{Theoretical}&
   &
   &
   \multicolumn{1}{c}{\checkmark}&
   \multicolumn{1}{c}{$\mathord{?}$}
   \\ \cline{3-8} 
 &
   &
  Redesigned protocol \cite{hall2015ieee}&
   \multicolumn{1}{c}{Experimental}&
   &
   \multicolumn{1}{c}{\checkmark}&
   &
   \multicolumn{1}{c}{\righthalfcircle}
   \\ \cline{3-8} 
 &
   &
  Transformer model \cite{bernabe2023detecting} &
   \multicolumn{1}{c}{Experimental}&
   \multicolumn{1}{c}{\checkmark}&
   \multicolumn{1}{c}{\checkmark}&
   &
   \multicolumn{1}{c}{\CIRCLE}
   \\ \cline{3-8} 
 &
   &
  Conceptual framework \cite{iphar2015detection}&
   \multicolumn{1}{c}{Theoretical}&
   &
   &
   \multicolumn{1}{c}{\checkmark}&
   \multicolumn{1}{c}{$\mathord{?}$}
   \\ \cline{3-8} 
 &
   &
  DSIAS and MTTPS \cite{su2017privacy} &
   \multicolumn{1}{c}{Theoretical}&
   \multicolumn{1}{c}{\checkmark}&
   &
   &
   \multicolumn{1}{c}{\Circle}
   \\ \cline{3-8} 
 &
   &
  Architectural solution \cite{kontopoulos2018countering} &
   \multicolumn{1}{c}{Experimental}&
   &
   \multicolumn{1}{c}{\checkmark}&
   &
   \multicolumn{1}{c}{\righthalfcircle}
   \\ \cline{3-8} 
 &
   &
  mIBC \cite{goudossis2019towards}\cite{goudosis2020secure}&
   \multicolumn{1}{c}{Theoretical}&
   &
   &
   \multicolumn{1}{c}{\checkmark}&
   \multicolumn{1}{c}{$\mathord{?}$}
   \\ \cline{3-8} 
 &
   &
  Authenticator module \cite{nguyen2020hardening} &
   \multicolumn{1}{c}{Experimental}&
   \multicolumn{1}{c}{\checkmark}&
   \multicolumn{1}{c}{\checkmark}&
   &
   \multicolumn{1}{c}{\lefthalfcircle}
   \\ \cline{3-8} 
 &
   &
  Auth-AIS protocol \cite{sciancalepore2021auth}&
   \multicolumn{1}{c}{Experimental}&
   \multicolumn{1}{c}{\checkmark}&
   \multicolumn{1}{c}{\checkmark}&
   &
   \multicolumn{1}{c}{\righthalfcircle}
   \\ \cline{3-8} 
 &
   &
  Detection algorithm \cite{kelly2022novel} &
   \multicolumn{1}{c}{Experimental}&
   \multicolumn{1}{c}{\checkmark}&
   &
   &\multicolumn{1}{c}{\LEFTcircle}
   \\ \cline{3-8} 
 &
   &
  Transformer model \cite{deng2023lightweight} &
   \multicolumn{1}{c}{Experimental}&
   \multicolumn{1}{c}{\checkmark}&
   \multicolumn{1}{c}{\checkmark}&
   &\multicolumn{1}{c}{\CIRCLE}
   \\ \hline
\multirow{7}{*}{GNSS / GPS} &
  \multirow{7}{*}{\begin{tabular}[c]{@{}l@{}}Intentional Jamming\\ Interference\\ Localised Jamming\\ Wide-area Jamming\\ GPS Service Denial\\ GPS Spoofing\end{tabular}} &
  Backup GPS \cite{grant2009gps} &
   \multicolumn{1}{c}{Experimental}&
   \multicolumn{1}{c}{\checkmark}&
   &
   &\multicolumn{1}{c}{\LEFTcircle}
   \\ \cline{3-8} 
 &
   &
  Detection framework \cite{bhatti2017hostile}&
   \multicolumn{1}{c}{Experimental}&
   \multicolumn{1}{c}{\checkmark}&
   &
   &\multicolumn{1}{c}{\LEFTcircle}
   \\ \cline{3-8} 
 &
   &
  Improved equipment \cite{drumhiller2017issues}&
   \multicolumn{1}{c}{Theoretical}&
   &
   &
   \multicolumn{1}{c}{\checkmark}&\multicolumn{1}{c}{$\mathord{?}$}
   \\ \cline{3-8} 
 &
   &
   Multi-sensor fusion \cite{medina2019gnss}&
   \multicolumn{1}{c}{Theoretical}&
   &
   &
   \multicolumn{1}{c}{\checkmark}&\multicolumn{1}{c}{$\mathord{?}$}
   \\ \cline{3-8} 
 &
   &
  Detection technique \cite{oligeri2020gnss}&
   \multicolumn{1}{c}{Experimental}&
   \multicolumn{1}{c}{\checkmark}&
   &
   &\multicolumn{1}{c}{\righthalfcircle}
   \\ \cline{3-8} 
 &
   &
  MANA framework \cite{spravil2023detecting}&
   \multicolumn{1}{c}{Experimental}&
   &
   \multicolumn{1}{c}{\checkmark}&
   &\multicolumn{1}{c}{\righthalfcircle}
   \\ \cline{3-8} 
 &
   &
  Detection system \cite{boudehenn2021navigation} &
   \multicolumn{1}{c}{Experimental}&
   \multicolumn{1}{c}{\checkmark}&
   \multicolumn{1}{c}{\checkmark}&
   &\multicolumn{1}{c}{\CIRCLE}
   \\ \hline
\multirow{6}{*}{ECDIS} &
  \multirow{6}{*}{\begin{tabular}[c]{@{}l@{}}File Manipulation\\ Directory Traversal \\ Outdated Software\\ Malware Introduction\\ HTTP Header Injection\\ Remote Code Execution\end{tabular}} &
  Update system \cite{ncc}&
   \multicolumn{1}{c}{Theoretical}&
   &
   &
   \multicolumn{1}{c}{\checkmark}&\multicolumn{1}{c}{$\mathord{?}$}
   \\ \cline{3-8} 
 &
   &
  Disable SMB v1 service \cite{svilicic2019raising}&
   \multicolumn{1}{c}{Theoretical}&
   &
   &
   \multicolumn{1}{c}{\checkmark}&\multicolumn{1}{c}{$\mathord{?}$}
   \\ \cline{3-8} 
 &
   &
  Backup arrangement \cite{svilicic2019assessing}  &
   \multicolumn{1}{c}{Theoretical}&
   &
   &
   \multicolumn{1}{c}{\checkmark}&\multicolumn{1}{c}{$\mathord{?}$}
   \\ \cline{3-8} 
 &
   &
  Proper configuration \cite{svilicic2019maritime}&
   \multicolumn{1}{c}{Theoretical}&
   &
   &
   \multicolumn{1}{c}{\checkmark}&\multicolumn{1}{c}{$\mathord{?}$}
   \\ \cline{3-8} 
 &
   &
  Access control \cite{svilicic2020paperless}  &
   \multicolumn{1}{c}{Theoretical}&
   &
   &
   \multicolumn{1}{c}{\checkmark}&\multicolumn{1}{c}{$\mathord{?}$}
   \\ \cline{3-8} 
 &
   &
  Secure setup \cite{svilicic2019shipboard} &
   \multicolumn{1}{c}{Theoretical}&
   &
   &
   \multicolumn{1}{c}{\checkmark}&\multicolumn{1}{c}{$\mathord{?}$}
   \\ \hline
RADAR &
  Ship Trajectory Attack&  Network monitoring \cite{longo2023attacking}&
   \multicolumn{1}{c}{Experimental}&
   \multicolumn{1}{c}{\checkmark}&
   \multicolumn{1}{c}{\checkmark}&
   &
   \multicolumn{1}{c}{\CIRCLE}\\ \hline
\multirow{2}{*}{VDR} &
  \multirow{2}{*}{\begin{tabular}[c]{@{}l@{}}Ransomware\\ Tampering\end{tabular}} &
  Network segmentation \cite{soner2023cybersecurity}&
   \multicolumn{1}{c}{Theoretical}&
   &
   &
   \multicolumn{1}{c}{\checkmark}&\multicolumn{1}{c}{$\mathord{?}$}
   \\ \cline{3-8} 
 &
   &
  Physical measures \cite{hopcraft2023raising} &
   \multicolumn{1}{c}{Theoretical}&
   &
   &
   \multicolumn{1}{c}{\checkmark}&\multicolumn{1}{c}{$\mathord{?}$}
   \\ \hline
\multirow{2}{*}{VSAT} &
  \multirow{2}{*}{\begin{tabular}[c]{@{}l@{}}Eavesdropping\\Spoofing\end{tabular}} &
  Encryption method \cite{pavur2020tale}&
   \multicolumn{1}{c}{Theoretical}&
   &
   &
   \multicolumn{1}{c}{\checkmark}&\multicolumn{1}{c}{$\mathord{?}$}
   \\ \cline{3-8} 
 &
   &
  lite-CA \cite{wu2018approach} &
   \multicolumn{1}{c}{Experimental}&
   &
   \multicolumn{1}{c}{\checkmark}&
   &\multicolumn{1}{c}{\righthalfcircle}
   \\ \hline
\multirow{2}{*}{GMDSS} &
  \multirow{2}{*}{\begin{tabular}[c]{@{}l@{}}DoS\\ Alert Disruption\end{tabular}} &
  Improved quality \cite{korcz2023key} &
   \multicolumn{1}{c}{Theoretical}&\multicolumn{1}{c}{\checkmark}
   &
   &
   &\multicolumn{1}{c}{\Circle}
   \\ \cline{3-8} 
 &
   &
  Global VTS system \cite{martinez2021global}&\multicolumn{1}{c}{Theoretical}
&\multicolumn{1}{c}{\checkmark}
   &
   &
   &\multicolumn{1}{c}{\Circle}
   \\ \bottomrule

\end{tabular}
\end{table*}

\section{Challenge Discussions}
\label{sec:challenges}

Maritime environments present unique challenges that are often overlooked in current cybersecurity literature. This section highlights several lesser-discussed issues that significantly impact the maritime sector's cybersecurity landscape.
\subsection{Device heterogeneity}
Maritime devices operate in a wide range of environments, ranging from ports and coastal areas and vast expanses of open seas. The diversity of maritime environments requires a corresponding diversity in device types. While this heterogeneity presents challenges for manufacturers and operators alike – including issues related to standardization and interoperability – it also ensures that appropriate technologies can be effectively deployed across various maritime scenarios. The maritime industry is facing an increasing demand for improved operational visibility at sea \cite{wang2020machine}. Enhanced operational visibility at sea contributes to overall safety by providing early detection of anomalies or hazards that could pose a threat to crew members or the environment. Maritime IoT addresses this requirement by making it possible to monitor vital indicators like fuel usage, machinery performance, and general vessel operations in real time. As a result, maritime stakeholders are able to make more informed decisions, optimize resource utilization, and improve overall efficiency and safety. However, the implementation of maritime IoT presents complexities. The system must accommodate a wide range of machine-type communication devices, from low-cost units with limited functionality to high-end devices offering advanced features \cite{wang2020machine}. Low-cost devices such as sensors and buoys often operate under power and energy constraints, posing challenges in selecting appropriate communication technologies and protocols that prioritize energy efficiency and longevity over high data rates or complex functionalities \cite{wang2020machine}. To address these challenges posed by device heterogeneity, edge computing and cloud-based solutions have been proposed \cite{dhivvya2017towards}. While cloud-based solutions and edge computing offer valuable tools for managing device heterogeneity in maritime IoT, cloud solutions require reliable internet connections, which may not always be available in open seas where face connectivity issues \cite{kim2019hierarchical}. In addition, implementing both cloud and edge solutions adds another layer of complexity to an already complex system of heterogeneous devices. 

\subsection{A culture of secrecy}
The maritime industry grapples with a unique challenge regarding cybersecurity incident transparency and information sharing. In stark contrast to the aviation sector, which has cultivated a culture of open reporting and collaborative information exchange on safety and security matters, the maritime domain often shrouds cyber incidents in secrecy, primarily due to concerns in reputational damage \cite{nimmich2007maritime}. In the United States alone, at least eighteen federal agencies have responsibility for regulating various aspects of maritime transportation, with little to no formal methods of coordinating their efforts \cite{national2004marine}. This regulatory fragmentation contributes to the "highly fragmented" and "near chaotic" nature of the maritime domain \cite{national2004marine}. This lack of coordination potentially discourages transparent reporting, and this may hinder the industry's collective ability to learn from past incidents and implement effective preventive measures against future cyber threats. Moreover, the opaque nature of maritime operations and regulations poses additional obstacles. For instance, it impedes the development of crucial technologies such as autonomous collision avoidance systems \cite{van2023towards}, thereby stunting the overall advancement of maritime technology. Overcoming this ingrained secrecy to foster greater transparency and information sharing will require significant shifts in industry practices, regulations, and mindsets. Key stakeholders must recognize that the benefits of shared knowledge and collaborative security efforts far outweigh the perceived risks of disclosure. True security lies not in secrecy, but in transparency. 

\subsection{Rapid technology evolution v.s. Slow-paced industry}
The slow pace of technological adoption in the global shipping fleet is evident in the statistics: only 0.3\% of operating vessels worldwide have implemented alternative energy solutions, with this figure rising slightly to 6.05\% for ships currently on order \cite{stalmokaite2020dynamic} This sluggish uptake is largely due to the long operational lifespan of vessels, typically 25-30 years, which creates significant inertia in the industry. Upgrading existing ships or investing in new builds with advanced technologies requires substantial capital, making fleet-wide changes extremely costly and time-consuming \cite{stalmokaite2023revival}. The capital-intensive nature of these investments leads to a lock-in effect, where shipping companies are hesitant to abandon existing technologies before fully deprecating their assets. This economic reality is compounded by the slow pace of regulatory change in the industry. For instance, despite recognizing the need to address greenhouse gas (GHG) emissions in the 1990s, the IMO did not adopt its first strategy for GHG reduction until 2018 \cite{stalmokaite2020dynamic}. The combination of long vessel lifespans, high capital costs, and a slow-moving regulatory environment creates a gap between the rapid evolution of new technologies and their implementation in the shipping industry.

\subsection{Limitations of technological solutions}

As previously discussed, while technologies such as blockchain, machine learning, and artificial intelligence show promise in addressing maritime cybersecurity challenges, they are not panaceas. Each technology comes with its own set of limitations, including scalability issues, susceptibility to new forms of attacks, and integration difficulties with existing systems. For example, the global scale and complexity of maritime supply chains strain blockchain's scalability and processing speed, potentially slowing down operations in an industry where time is critical \cite{li2020blockchain}.  It is essential for the maritime industry to approach these technologies carefully, recognizing both their potential benefits and inherent constraints. Additionally, rather than relying solely on a single advanced technology, a more effective strategy involves implementing a multi-layered technological approach. This may entail combining various technologies to establish a stronger defense. For instance, utilizing AI for threat detection, blockchain for secure data sharing, and quantum-resistant cryptography for future-proofing communications. This layered approach can help mitigate the weaknesses of individual technologies. Overall, while advanced technologies hold promise for enhancing cybersecurity in the maritime sector, it is crucial to carefully evaluate their capabilities and limitations before implementation.

\section{Future Directions}
\label{sec:direction}

While the maritime industry's future encompasses numerous potential directions, this section focuses on current trending topics and areas undergoing active implementation. These selected future directions are expected to significantly reshape the maritime security landscape. 
\subsection{Maritime Cloud}
The maritime cloud concept is currently in its formative stages, with ongoing projects and evolving implementations shaping its development.  Park \cite{themaritimecloud} mentioned that the maritime cloud is closely tied to the development of e-Navigation, a strategic initiative spearheaded by the IMO to harmonize and enhance navigation systems globally. The integration of the maritime cloud with e-Navigation strategies could offer a opportunity to facilitate seamless information exchange across diverse maritime stakeholders. It has the potential to transcend the limitations of specific communication systems or channels, creating a unified platform that connects vessels, ports, coastal authorities and other maritime actors. Cloud technology offers numerous other advantages in the maritime domain, including enhanced data management \cite{cristea2017operational}, improved maritime safety and security with distributed smart surveillance systems \cite{lee2017geocasting}, cost reduction \cite{ristov2014implemetation}, system compatibility \cite{themaritimecloud}, and real-time monitoring of vessel positions and status \cite{joszczuk2012benefits}. Despite these promising applications, further research is imperative across multiple domain within maritime cloud computing, such as security concerns in cloud implementation \cite{botunac2017analysis}.

\subsection{Testbeds}
Maritime-specific cybersecurity testbeds can be used to simulate cyber attacks on ship systems, allowing for the development and testing of defense mechanisms without risking actual vessels or infrastructure. In response to the growing need for advanced testing and research capabilities in the maritime sector, several notable testbeds have been established worldwide. The eMaritime Integrated Reference Platform (eMIR) \cite{hahn2015simulation}, developed in Germany, combines both virtual and physical components for providing a robust environment to test maritime cyber-physical systems. In the United Kingdom, the University of Plymouth hosts the Cyber-SHIP Lab \cite{tam2019cyber}, a specialised facility for researchers to carry out cyber attack experiments. Meanwhile, in Singapore, the Singapore Polytechnic houses the Advanced Navigation Research Simulator (ANRS) \cite{cems}, which is designed to support maritime training and education. Existing testbeds rely heavily on simulation rather than real maritime hardware, as physical testbeds incorporating actual maritime equipment are costly to construct and maintain \cite{yamin2020cyber}\cite{zhou2024need}. Each testbed architecture has its distinct advantages and limitations. Virtual testbeds enable testing of potentially dangerous scenarios without risking damage to physical equipment or endangering personnel \cite{longo2023macyste}; however, they often struggle to accurately replicate wireless communications \cite{amro2021communication}, which are crucial in maritime settings. There is a need for a balance between virtual and physical components to achieve greater realism in maritime testbeds. While current testbeds are valuable, we believe they require continued investment and innovation to better support the evolving needs of the maritime industry.

\section{Conclusion}
\label{sec:conclusion}

This review underscores the complexity of securing maritime systems and emphasized the importance of a multi-faceted approach that combines technological innovation, regulatory frameworks, and cultural shifts within the industry. While promising solutions like AI-driven threat detection and blockchain-based authentication offer new defensive capabilities, they must be implemented thoughtfully to address the unique challenges of the maritime environment. Moving forward, the development of maritime-specific cybersecurity testbeds, the evolution of the maritime cloud, and increased transparency in incident reporting will be crucial in building a more secure and resilient global maritime infrastructure. As the industry continues to navigate these digital waters, collaboration between stakeholders, adaptive regulatory frameworks, and continuous research will be essential in safeguarding the future of maritime operations against ever-evolving cyber threats.

\section*{Acknowledgments}
We would like to express our gratitude to Bureau Veritas and ClassNK for their contributions to this survey paper. Their dedication and professional insights have been crucial to the successful completion of this study. We extend particular appreciation to Arthur Piqu{\'e}e-Audrain from the Bureau Veritas Marine \& Offshore cybersecurity section, whose thorough analysis and expertise in maritime cybersecurity have added substantial value to our research findings. We are equally grateful to Kaoru Shimamoto from ClassNK, whose comprehensive insights have strengthened the depth of our study. The authors declare no conflict of interest. This research is supported by the National Research Foundation, Singapore (NRF), Maritime and Port Authority of Singapore (MPA) and Singapore Maritime Institute (SMI) under its Maritime Transformation Programme (Project No. SMI-2022-MTP-05). Any opinions, findings and conclusions or recommendations expressed in this material are those of the author(s) and do not reflect the views of NRF, MPA and SMI.

\appendix

\section{APPENDIX}
In this section, we have included supplementary information for our survey. Firstly, we present definitions for a list of discussed vessel components. We further analyzed their vulnerabilities, highlighting the possible attacks that can happen. 

\subsection{Definitions of Vessel Components}
Vessel components refer to the various systems, equipment, and structural elements that collectively form a ship or marine vessel. Modern vessels incorporate a range of digital systems and networks that control and monitor these physical components, including integrated bridge systems, engine control and monitoring systems, and onboard networks. In this survey, we mainly focus on the components found in IBS because those components are mostly discussed. 

\textbf{AIS:}
The AIS is a Very High Frequency (VHF) radio broadcasting system that is used for traffic monitoring (Vessel Traffic Services), collision avoidance, search and rescue operations, aids to navigation, weather forecast (Australian Maritime Safety Authority) and accident investigation \cite{bothur2017critical}\cite{rivkin2023maritime}.

\textbf{GNSS:}
GNSS is an umbrella term for satellite-based navigation systems that provide global coverage for position, velocity, and timing information. There are currently four primary GNSS in operation: GPS operated by the United States, GLONASS operated by Russia, Galileo operated by the European Union, and BeiDou operated by China. In the maritime context, GNSS plays a crucial role in various aspects of operations, including navigation, vessel monitoring, autonomous shipping, timing, and integration with other onboard systems.

\textbf{VDR:}
VDR is an essential system that acts as a "black box" to record key parametric data about a ship's voyage, which can then be recovered and analyzed to investigate the causes of a maritime accident. VDR continuously stores data related to the status, command, and control of the ship from various sensors and systems, including date/time, ship's position, speed, heading, bridge audio, radio communications, radar data, hull openings status, watertight doors status, etc \cite{piccinelli2013modern}.

\textbf{RADAR:}
RADAR is an object detection system that emits radio waves and analyzes the reflected signals to detect the presence, direction, and range of objects in the surrounding environment \cite{awan2019understanding}. It is indispensable for maintaining situational awareness and ensuring the safety of maritime operations, particularly in low visibility conditions or when navigating through congested waters. 

\textbf{ECDIS:}
ECDIS is a computer-based navigation system that serves as a digital alternative to traditional paper nautical charts, providing navigators with real-time, interactive, and comprehensive information about the ship's position, course, and surrounding environment. It combines information from multiple navigational aids and sensors, including GNSS, Radar, and AIS \cite{rivkin2023maritime}. This information is superimposed on digital charts and empowers navigation officers to make informed decisions and maintain situational awareness.

\textbf{VSAT:}
Maritime VSAT systems provide essential connectivity and communication services, including internet access for crew and passengers, operational communications between ship and shore, real-time navigation updates, weather forecasts, and safety information. 

\textbf{GMDSS:}
The GMDSS was developed by the International Maritime Organization (IMO) and became fully operational in 1999. It is designed to perform the main functions such as transmitting ship-to-shore distress alerts, coordinating search and rescue operations and providing general radiocommunications \cite{ilcev2020new}. The implementation of GMDSS has significantly improved maritime safety and rescue operations globally, providing a standardized system for distress communication and alerting across international waters.
\subsection{Vulnerabilities of Vessel Components}
The vulnerabilities of vessel components encompass a wide range of potential susceptibilities that could be exploited to compromise the safety, security, or operational integrity of a ship. These vulnerabilities can arise from various sources, including inherent design flaws, manufacturing defects, inadequate maintenance, human error, or intentional tampering.  

\textbf{Vulnerability of AIS:}
It was developed in the 1990s with a primary focus on enhancing maritime safety and navigation, but its design did not thoroughly consider the potential risks posed by active cyber attacks or malicious actors seeking to exploit vulnerabilities in the system. Numerous known security flaws exist \cite{goudossis2019towards}\cite{kessler2020maritime}:

\begin{itemize}
  \item The absence of built-in mechanisms in AIS to authenticate the geographical origin of transmitted messages allows a device to broadcast AIS data that is not accurate in indicating its location. Because of this flaw, a hacker might pretend to be an AIS transmitter and fool other ships or shore stations about the actual location of the transmitter.
  \item The absence of timestamp information in AIS communications exposes the system to replay assaults. Attackers have the capacity to record and replay valid AIS broadcasts, which compromises the system's dependability for forensic investigation and situational awareness by giving misleading perceptions about a vessel's previous position.
  \item AIS lacks message authentication, the system is unable to confirm the sender's actual identity. Anyone with the ability to send AIS packets can impersonate any other AIS device, which puts vessels that depend on the accuracy of AIS messages at danger for confusion, deceit, and accidents.
  \item Since AIS does not have mechanisms to guarantee the integrity of sent data, it is impossible to confirm that the data sent by an AIS device is accurate and has not been tampered with. 
  \item AIS-broadcast communications lack encryption. Openly available AIS data may be found on a number of websites, including Marine Traffic, and can offer a wealth of information about ships and their travels, including specifics about the ship's identity, cargo, and current position \cite{bothur2017critical}\cite{hall2015ieee}. 
\end{itemize}

\textbf{Vulnerability of GNSS:}
The main GNSS vulnerabilities that have been thoroughly examined are the radio frequency hazards associated with jamming and spoofing. Extreme space weather events have been identified as a significant vulnerability for GNSS, in addition to radio frequency concerns. In contrast to physical assaults, which are mostly theoretical because mature nation-states demand highly developed capabilities, extreme space weather phenomena represent a genuine and continuous threat to GNSS satellites and their communications. These occurrences are real threats, even though their frequency and intensity vary over the course of the 11-year solar cycle \cite{owens2021extreme}. There are several reasons why GNSS are susceptible \cite{drumhiller2017issues}\cite{10.1145/2897166}:

\begin{itemize}
  \item By the time GNSS signals reach the Earth's surface, they are inherently weak because they are coming from satellites in a medium Earth orbit. These signals are weaker when they reach GNSS receivers due to the great distances they must travel and the power constraints of satellite transmitters. Due to this intrinsic weakness, hostile actors can use readily available jamming equipment to disrupt or block GNSS signals.
  \item Many GNSS signals are open and unencrypted, which makes them vulnerable to spoofing attacks, in which malevolent actors create and disseminate phony signals that appear to be real. Meaconing is a kind of GNSS spoofing attack in which the attacker first adds a predetermined delay to the GNSS signals they have intercepted, then rebroadcasts them. 

  \item GNSS has a significant reliance on satellites, which are vulnerable to potential physical attacks (such as those using anti-satellite weapons) as well as space weather phenomena (such solar flares). 
\end{itemize}

\textbf{Vulnerability of VDR:}
VDR data could be contested in court and declared inadmissible as evidence if it turns out to be easily altered or manipulated. Legal proceedings pertaining to maritime accidents, insurance claims, and liability issues may be significantly impacted by this \cite{ashour2013influence}. Consequently, VDR may be exposed to hazards associated with data availability, data integrity problems, and confidentiality breaches.

\textbf{Vulnerability of RADAR:}
A number of subsystems, including antennas, transmitters, receivers, processors, and displays, make up the intricate marine RADAR systems \cite{cohen2019security}. There could be vulnerabilities in every component, which would increase the number of attack surfaces. However, given the specialized nature of marine radar systems and the challenges associated with conducting thorough security assessments in real-world scenarios, there might be fewer opportunities to thoroughly inspect the system for vulnerabilities. The task of simulating actual circumstances for security testing is hampered by the complicated operational settings, which include the sea environment, weather, and the requirement for uninterrupted functionality.

\textbf{Vulnerability of ECDIS:}
The vulnerabilities in ECDIS are largely caused by a combination of technological factors and organisational issues as the maritime industry has been slow to adapt to the rapidly evolving cyber threat landscape. The main vulnerabilities identified in ECDIS are \cite{svilicic2019raising}\cite{svilicic2019assessing}: 

\begin{itemize}
  \item ECDIS often operates on outdated operating systems like Windows XP or Windows 7, which no longer receive security updates. Additionally, third-party applications such as web servers and remote desktop tools are frequently outdated and contain known vulnerabilities.
  
  \item Security patches for ECDIS systems are not consistently applied as it is a manual and time-consuming process.
  \item Insecure network configuration and services, such as directory traversal, unsafe HTTP methods allowed, and header injection vulnerabilities in the Apache web server \cite{ncc}.
  \item Identical hardware/software used for primary and backup ECDIS.

\end{itemize}

\textbf{Vulnerability of VSAT:}
Many maritime operators appear to be unaware of the risks associated with transmitting sensitive data over satellite links without adequate security measures \cite{wu2018approach}, and numerous maritime VSAT networks continue to use outdated and insecure protocols \cite{pavur2020tale}. The author \cite{santamarta2018last} observes a trend where historically isolated devices, such as VSAT systems, are now being designed with additional communication technologies like WiFi and Bluetooth. 

\textbf{Vulnerability of GMDSS:}
The radio and satellite communications used by GMDSS can be intentionally jammed or unintentionally interfered with, which could disrupt its functionalities \cite{tam2018maritime} GMDSS relies on various interconnected systems. A vulnerability in one system could potentially affect the entire GMDSS functionality. Systems like GPS, which GMDSS relies on for positioning, can be spoofed to provide false location data \cite{ilcev2020new}.

\subsection{Acronyms}
Lastly, we have provided a list of all the acronyms (See Table \ref{tab:acronyms}) used in this survey.
\begin{table*}[h]\footnotesize
\centering
  \caption{Acronyms}
  \label{tab:acronyms}
  \begin{tabular}{c|c|c|c}
    \toprule
    Acronym &Description&Acronym &Description \\
    \toprule
    ABS&American Bureau of Shipping&IMO&International Maritime Organisation\\
    
    \midrule
    ADS-B&Automatic Dependent Surveillance-Broadcast&IoT&Internet of Things\\
    \midrule
    AI-PAML&Adaptive Incremental Passive-Aggressive Machine Learning&ISPS&International Ship and Port Facility Security\\
    \midrule
    AIS&Automatic Identification System&IT&Information Technology\\
    \midrule
    ANRS&Advanced Navigation Research Simulator&lite-CA&Lite Certification Authority\\
    \midrule
    BIMCO&Baltic and International Maritime Council&LSTM&Long Short-Term Memory\\
    \midrule
    DAU&Data Acquisition Unit&MARL&Multi-Agent Reinforcement Learning\\
    \midrule
    DDoS&Distributed Denial of Service&MCAD&Maritime Cyber Attack Database\\
    \midrule
    DoS&Denial of Service&MTPPS&Mix-zone Based Trajectory Privacy Protection Scheme\\
    \midrule
    DRFM&Digital Radio Frequency Memory&NAVTEX&Navigation Telex\\
    \midrule
    DSIAS&Digital Signature Based Identity Authentication Scheme&NIST&National Institute of Standards and Technology\\
    \midrule
    EA&Electronic Attack&NMEA&National Marine Electronics Association\\
    \midrule
    ECDIS&Electronic Chart Display and Information System&OT & Operational Technology\\
    \midrule
    ECDSA&Elliptic Curve Digital Signature Algorithm&PDCA&Plan-Do-Check-Act\\
    \midrule
    eMIR&eMaritime Integrated Reference Platform&PKI&Public Key Infrastructure\\
    \midrule
    ENC&Electronic Navigational Chart&PLA&Port of London Authority\\
    \midrule
    FMEA&Failure Mode and Effect Analysis&PLCs&Programmable Logic Controllers\\
    \midrule
    FMECA&Failure Mode, Effect, and Criticality Analysis&PoC&Proof-of-Concept\\
    \midrule
    GHG&Greenhouse Gas&PPI&Plan Position Indicator\\
    \midrule
    GP&Gaussian Process&RSSI&Received Signal Strength Indicator\\
    \midrule
    GIS&Geographic Information System&S2ML&Dual Stack Machine Learning\\
    \midrule
    GMDSS&Global Maritime Distress and Safety System&SATCOM&Satellite Communications\\
    \midrule
    GNSS&Global Navigation Satellite System&SEI&Specific Emitter Identification\\
    \midrule
    GPS&Global Positioning System&RADAR&RAdio Detection And Ranging\\
    \midrule
    HAZOP&Hazard and Operability Analysis&UAV&Unmanned Aerial Vehicle\\
    \midrule
    HFACS&Human Factors Analysis and Classification System&VDES&VHF Data Exchange System\\
    \midrule
    HW-F&High Weight Function&VDR&Voyage Data Recorder\\
    \midrule 
    IACS&International Association of Classification Society&VHF&Very High Frequency\\
    \midrule  
    IBS&Integrated Bridge System&VSAT&Very Small Aperture Terminal\\
    \midrule

  \end{tabular}
\end{table*}

\bibliographystyle{IEEEtran}
\bibliography{refs}
\clearpage

\section{Biography Section}
 

\begin{IEEEbiography}
    [{\includegraphics[width=1in,height=1.25in,clip,keepaspectratio]{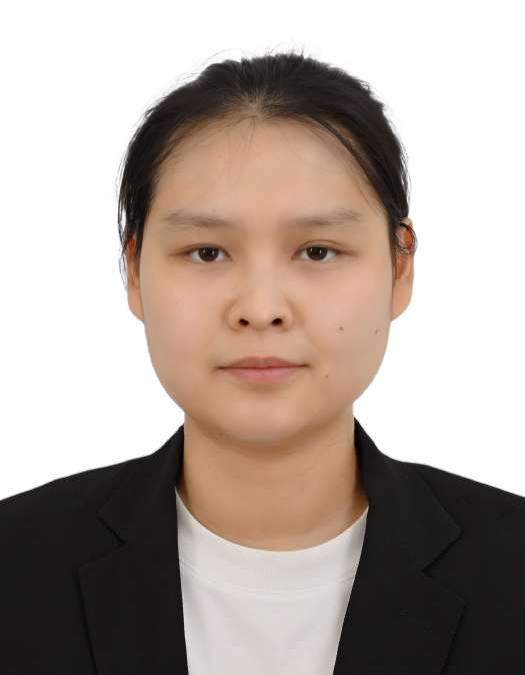}}]{Meixuan Li} received the B.E. degree (Hons.) in computer science from Singapore University of Technology and Design (SUTD), Singapore, in 2022. After graduation, she works as a Research Officer for iTrust at SUTD. Her work is primarily focused on the Maritime Testbed of Shipboard Operational Technology (MariOT) project, a strategic initiative funded through the National Satellite of Excellence (NSoE) program by the Singapore government. Her research interests include cyber-physcial security, operational technology system, maritime cybersecurity, and autonomous shipping.
\end{IEEEbiography}
\begin{IEEEbiography}
    [{\includegraphics[width=1in,height=1.25in,clip,keepaspectratio]{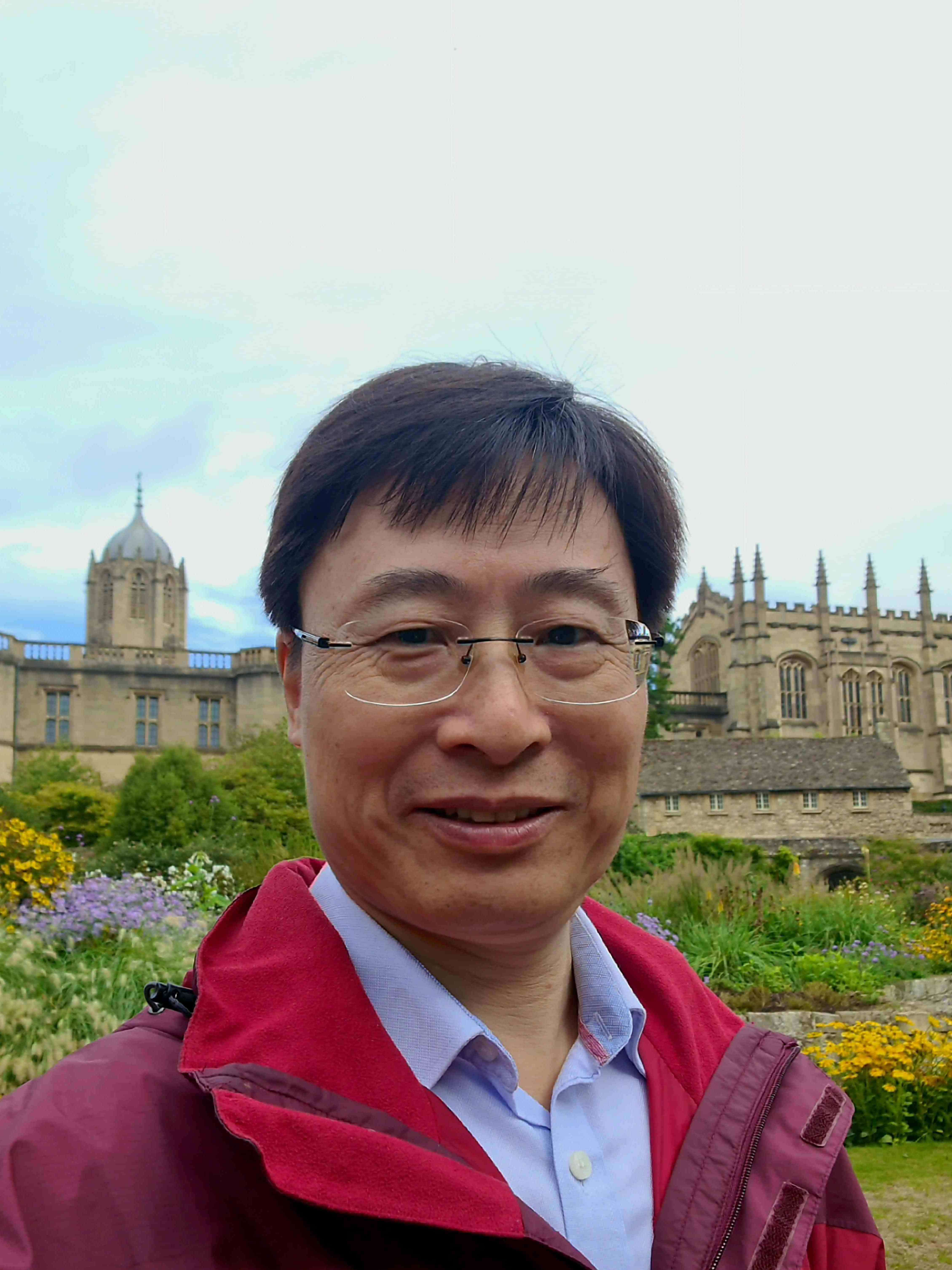}}]{Jianying Zhou} is a professor and center director for iTrust at Singapore University of Technology and Design (SUTD). He received PhD in Information Security from Royal Holloway, University of London. His research interests are in applied cryptography and network security, cyber-physical system security, mobile and wireless security. He has published papers at international conferences (IEEE S\&P, ACM CCS, USENIX Security, NDSS etc.) and journals (IEEE TIFS, IEEE TDSC etc.). He received ESORICS'15 Best Paper Award and ACSAC'23 Distinguished Paper Award. He is a co-founder \& steering committee co-chair of ACNS, steering committee chair of ACM AsiaCCS, and steering committee member of Asiacrypt. He is associate editor-in-chief of IEEE Security \& Privacy. He is also an ACM Distinguished Member. He received the ESORICS Outstanding Contribution Award in 2020, in recognition of his contributions to the community.
\end{IEEEbiography}
\begin{IEEEbiography}
    [{\includegraphics[width=1in,height=1.25in,clip,keepaspectratio]{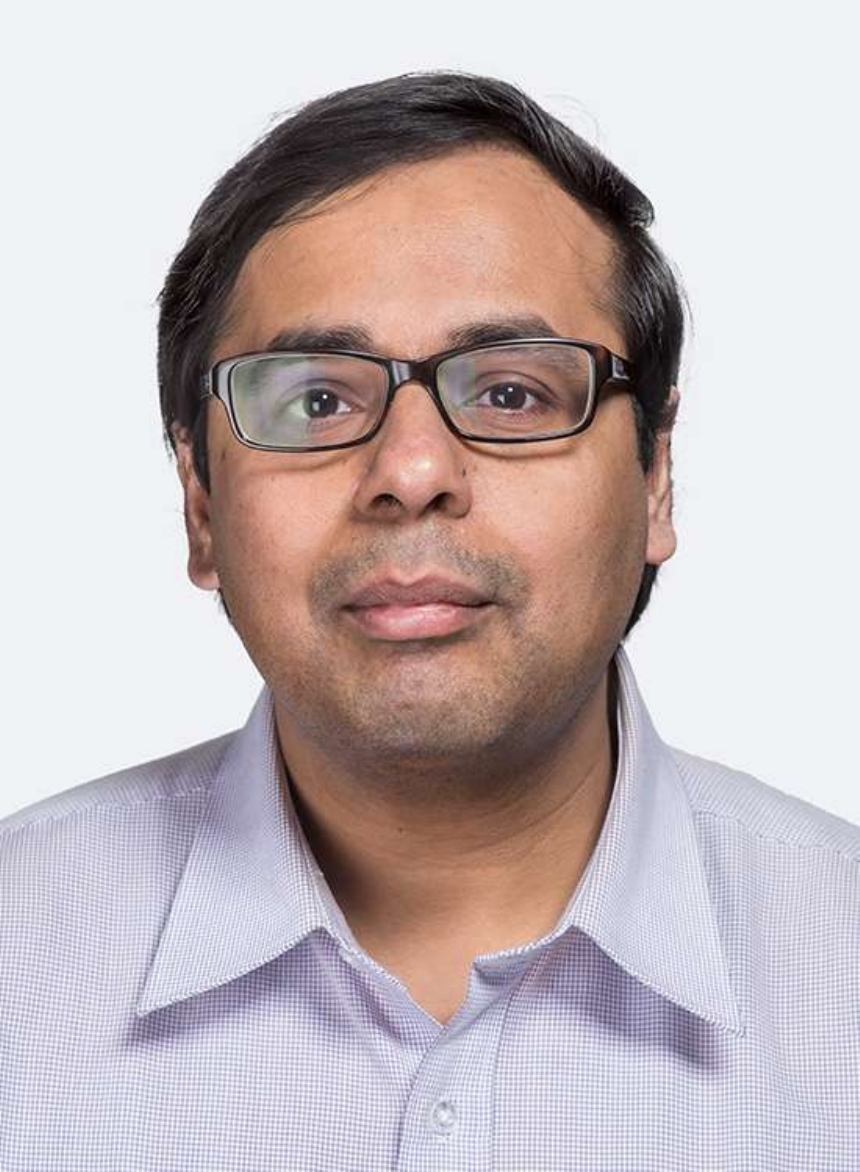}}]{Sudipta Chattopadhyay} received the PhD degree in computer science from the National University of Singapore, Singapore, in 2013. He is an associate professor with the Information Systems Technology and Design Pillar, Singapore University of Technology and Design, Singapore. In his doctoral dissertation, he researched on Execution-Time Predictability, focusing on Multicore Platforms. He seeks to understand the influence of execution platform on critical software properties, such as performance, energy, robustness, and security. His research interests include program analysis, embedded systems, and compilers. His serves in the review board of the IEEE Transactions on Software Engineering.
\end{IEEEbiography}
\begin{IEEEbiography}
    [{\includegraphics[width=1in,height=1.25in,clip,keepaspectratio]{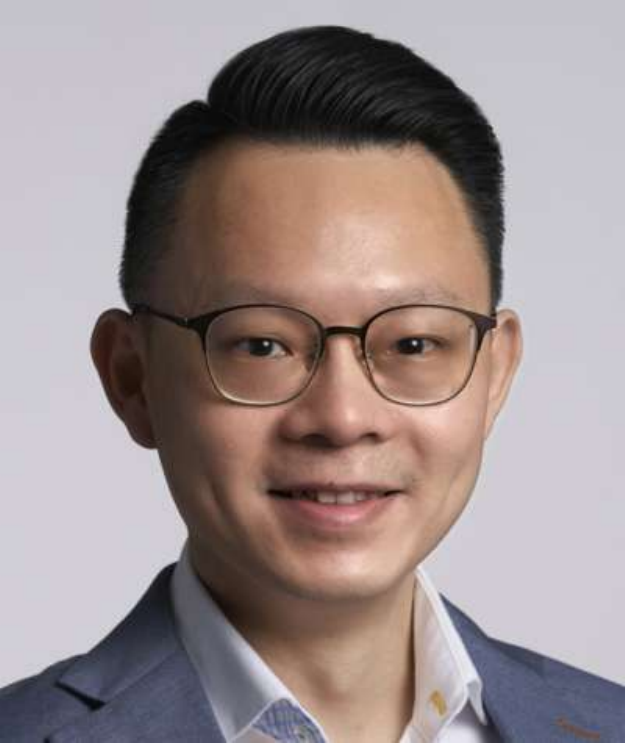}}]{Mark Goh} has a B.Eng in Civil Engineering (Hons.) and an M.Eng in Environmental Science and Engineering (Hons.) from the National University of Singapore (NUS). He started his career at the Maritime and Port Authority of Singapore (MPA), managing R\&D projects and programmes in shipboard technology, renewable energy and civil and structural engineering. Mark moved on to develop stakeholder engagement and partnership strategies for a diversity of stakeholders at the national level. At iTrust, he oversees international collaborations, office and lab administration, and business development, securing research funds and revenue for the Centre. 
\end{IEEEbiography}

\end{document}